\documentclass[twocolumn]{aastex62}

\shorttitle{Subaru High-$z$ Exploration of Low-Luminosity Quasars (SHELLQs) X}
\shortauthors{Matsuoka et al.}

\begin{document}

\title{Subaru High-{\scriptsize $z$} Exploration of Low-Luminosity Quasars (SHELLQ{\scriptsize s}). X. Discovery of 35 Quasars and Luminous Galaxies at $5.7 \le$ {\scriptsize $z$} $\le 7.0$}

\correspondingauthor{Yoshiki Matsuoka}
\email{yk.matsuoka@cosmos.ehime-u.ac.jp}

\author{Yoshiki Matsuoka}
\affil{Research Center for Space and Cosmic Evolution, Ehime University, Matsuyama, Ehime 790-8577, Japan.}

\author{Kazushi Iwasawa}
\affil{ICREA and Institut de Ci{\`e}ncies del Cosmos, Universitat de Barcelona, IEEC-UB, Mart{\'i} i Franqu{\`e}s, 1, 08028 Barcelona, Spain.}

\author{Masafusa Onoue}
\affil{Max Planck Institut f\"{u}r Astronomie, K\"{o}nigstuhl 17, D-69117, Heidelberg, Germany}

\author{Nobunari Kashikawa}
\affil{Department of Astronomy, School of Science, The University of Tokyo, Tokyo 113-0033, Japan.}
\affil{National Astronomical Observatory of Japan, Mitaka, Tokyo 181-8588, Japan.}
\affil{Department of Astronomical Science, Graduate University for Advanced Studies (SOKENDAI), Mitaka, Tokyo 181-8588, Japan.}

\author{Michael A. Strauss}
\affil{Princeton University Observatory, Peyton Hall, Princeton, NJ 08544, USA.}

\author{Chien-Hsiu Lee}
\affil{National Optical Astronomy Observatory, 950 North Cherry Avenue, Tucson, AZ 85719, USA.}

\author{Masatoshi Imanishi}
\affil{National Astronomical Observatory of Japan, Mitaka, Tokyo 181-8588, Japan.}
\affil{Department of Astronomical Science, Graduate University for Advanced Studies (SOKENDAI), Mitaka, Tokyo 181-8588, Japan.}

\author{Tohru Nagao}
\affil{Research Center for Space and Cosmic Evolution, Ehime University, Matsuyama, Ehime 790-8577, Japan.}

\author{Masayuki Akiyama}
\affil{Astronomical Institute, Tohoku University, Aoba, Sendai, 980-8578, Japan.}

\author{Naoko Asami}
\affil{Seisa University, Hakone-machi, Kanagawa, 250-0631, Japan.}

\author{James Bosch}
\affil{Princeton University Observatory, Peyton Hall, Princeton, NJ 08544, USA.}

\author{Hisanori Furusawa}
\affil{National Astronomical Observatory of Japan, Mitaka, Tokyo 181-8588, Japan.}

\author{Tomotsugu Goto}
\affil{Institute of Astronomy and Department of Physics, National Tsing Hua University, Hsinchu 30013, Taiwan.}

\author{James E. Gunn}
\affil{Princeton University Observatory, Peyton Hall, Princeton, NJ 08544, USA.}

\author{Yuichi Harikane}
\affil{Institute for Cosmic Ray Research, The University of Tokyo, Kashiwa, Chiba 277-8582, Japan}
\affil{Department of Physics, Graduate School of Science, The University of Tokyo, Bunkyo, Tokyo 113-0033, Japan}

\author{Hiroyuki Ikeda}
\affil{National Astronomical Observatory of Japan, Mitaka, Tokyo 181-8588, Japan.}

\author{Takuma Izumi}
\affil{National Astronomical Observatory of Japan, Mitaka, Tokyo 181-8588, Japan.}

\author{Toshihiro Kawaguchi}
\affil{Department of Economics, Management and Information Science, Onomichi City University, Onomichi, Hiroshima 722-8506, Japan.}

\author{Nanako Kato}
\affil{Graduate School of Science and Engineering, Ehime University, Matsuyama, Ehime 790-8577, Japan.}

\author{Satoshi Kikuta}
\affil{National Astronomical Observatory of Japan, Mitaka, Tokyo 181-8588, Japan.}
\affil{Department of Astronomical Science, Graduate University for Advanced Studies (SOKENDAI), Mitaka, Tokyo 181-8588, Japan.}

\author{Kotaro Kohno}
\affil{Institute of Astronomy, The University of Tokyo, Mitaka, Tokyo 181-0015, Japan.}
\affil{Research Center for the Early Universe, University of Tokyo, Tokyo 113-0033, Japan.}

\author{Yutaka Komiyama}
\affil{National Astronomical Observatory of Japan, Mitaka, Tokyo 181-8588, Japan.}
\affil{Department of Astronomical Science, Graduate University for Advanced Studies (SOKENDAI), Mitaka, Tokyo 181-8588, Japan.}

\author{Shuhei Koyama}
\affil{Research Center for Space and Cosmic Evolution, Ehime University, Matsuyama, Ehime 790-8577, Japan.}

\author{Robert H. Lupton}
\affil{Princeton University Observatory, Peyton Hall, Princeton, NJ 08544, USA.}

\author{Takeo Minezaki}
\affil{Institute of Astronomy, The University of Tokyo, Mitaka, Tokyo 181-0015, Japan.}

\author{Satoshi Miyazaki}
\affil{National Astronomical Observatory of Japan, Mitaka, Tokyo 181-8588, Japan.}
\affil{Department of Astronomical Science, Graduate University for Advanced Studies (SOKENDAI), Mitaka, Tokyo 181-8588, Japan.}


\author{Hitoshi Murayama}
\affil{Kavli Institute for the Physics and Mathematics of the Universe, WPI, The University of Tokyo,Kashiwa, Chiba 277-8583, Japan.}

\author{Mana Niida}
\affil{Graduate School of Science and Engineering, Ehime University, Matsuyama, Ehime 790-8577, Japan.}

\author{Atsushi J. Nishizawa}
\affil{Institute for Advanced Research, Nagoya University, Furo-cho, Chikusa-ku, Nagoya 464-8602, Japan.}

\author{Akatoki Noboriguchi}
\affil{Graduate School of Science and Engineering, Ehime University, Matsuyama, Ehime 790-8577, Japan.}

\author{Masamune Oguri}
\affil{Department of Physics, Graduate School of Science, The University of Tokyo, Bunkyo, Tokyo 113-0033, Japan}
\affil{Kavli Institute for the Physics and Mathematics of the Universe, WPI, The University of Tokyo,Kashiwa, Chiba 277-8583, Japan.}
\affil{Research Center for the Early Universe, University of Tokyo, Tokyo 113-0033, Japan.}

\author{Yoshiaki Ono}
\affil{Institute for Cosmic Ray Research, The University of Tokyo, Kashiwa, Chiba 277-8582, Japan}

\author{Masami Ouchi}
\affil{Institute for Cosmic Ray Research, The University of Tokyo, Kashiwa, Chiba 277-8582, Japan}
\affil{Kavli Institute for the Physics and Mathematics of the Universe, WPI, The University of Tokyo,Kashiwa, Chiba 277-8583, Japan.}

\author{Paul A. Price}
\affil{Princeton University Observatory, Peyton Hall, Princeton, NJ 08544, USA.}

\author{Hiroaki Sameshima}
\affil{Institute of Astronomy, The University of Tokyo, Mitaka, Tokyo 181-0015, Japan.}

\author{Andreas Schulze}
\affil{National Astronomical Observatory of Japan, Mitaka, Tokyo 181-8588, Japan.}


\author{John D. Silverman}
\affil{Kavli Institute for the Physics and Mathematics of the Universe, WPI, The University of Tokyo,Kashiwa, Chiba 277-8583, Japan.}

\author{Naoshi Sugiyama}
\affil{Kavli Institute for the Physics and Mathematics of the Universe, WPI, The University of Tokyo,Kashiwa, Chiba 277-8583, Japan.}
\affil{Graduate School of Science, Nagoya University, Furo-cho, Chikusa-ku, Nagoya 464-8602, Japan.}

\author{Philip J. Tait}
\affil{Subaru Telescope, Hilo, HI 96720, USA.}

\author{Masahiro Takada}
\affil{Kavli Institute for the Physics and Mathematics of the Universe, WPI, The University of Tokyo,Kashiwa, Chiba 277-8583, Japan.}

\author{Tadafumi Takata}
\affil{National Astronomical Observatory of Japan, Mitaka, Tokyo 181-8588, Japan.}
\affil{Department of Astronomical Science, Graduate University for Advanced Studies (SOKENDAI), Mitaka, Tokyo 181-8588, Japan.}

\author{Masayuki Tanaka}
\affil{National Astronomical Observatory of Japan, Mitaka, Tokyo 181-8588, Japan.}
\affil{Department of Astronomical Science, Graduate University for Advanced Studies (SOKENDAI), Mitaka, Tokyo 181-8588, Japan.}

\author{Ji-Jia Tang}
\affil{Institute of Astronomy and Astrophysics, Academia Sinica, Taipei, 10617, Taiwan.}

\author{Yoshiki Toba}
\affil{Department of Astronomy, Kyoto University, Sakyo-ku, Kyoto, Kyoto 606-8502, Japan.}

\author{Yousuke Utsumi}
\affil{Kavli Institute for Particle Astrophysics and Cosmology, Stanford University, CA 94025, USA.}

\author{Shiang-Yu Wang}
\affil{Institute of Astronomy and Astrophysics, Academia Sinica, Taipei, 10617, Taiwan.}

\author{Takuji Yamashita}
\affil{National Astronomical Observatory of Japan, Mitaka, Tokyo 181-8588, Japan.}



\begin{abstract}
We report the discovery of 28 quasars and 7 luminous galaxies at $5.7 \le z \le 7.0$.
This is the tenth in a series of papers from the Subaru High-$z$ Exploration of Low-Luminosity Quasars (SHELLQs) project, which 
exploits the deep multi-band imaging data produced by the Hyper Suprime-Cam (HSC) Subaru Strategic Program survey.
The total number of spectroscopically identified objects in SHELLQs has now grown to  
93 high-$z$ quasars, 31 high-$z$ luminous galaxies, 16 [\ion{O}{3}] emitters at $z \sim 0.8$, and 65 Galactic cool dwarfs (low-mass stars and brown dwarfs).
These objects were found over 900 deg$^2$, 
surveyed by HSC between 2014 March and 2018 January.
The full quasar sample includes 18 objects with very strong and narrow Ly$\alpha$ emission, whose stacked spectrum is clearly different from that of other quasars or galaxies.
While the stacked spectrum shows \ion{N}{5} $\lambda$1240 emission and resembles that of lower-$z$ narrow-line quasars, the small Ly$\alpha$ width may suggest
a significant contribution from the host galaxies.
Thus these objects may be composites of quasars and star-forming galaxies.
\end{abstract}

\keywords{dark ages, reionization, first stars --- galaxies: active --- galaxies: high-redshift --- intergalactic medium --- quasars: general --- quasars: supermassive black holes}

\section{Introduction} \label{sec:intro}

Quasars in the high-$z$ universe\footnote{
Throughout this paper, ``high-$z$" denotes $z > 5.7$, where quasars are observed as $i$-band dropouts in the 
Sloan Digital Sky Survey (SDSS) filter system \citep{fukugita96}.
The term ``$X$-band dropouts" or ``$X$-dropouts" refers to objects which are much fainter (and often invisible) in the $X$ and bluer bands 
than in the redder bands.}
have been used as an unique probe into early cosmic history. 
The progress of cosmic reionization can be measured from \ion{H}{1} absorption imprinted on the rest-frame ultraviolet (UV) spectrum of a high-$z$ quasar; this absorption is very sensitive to the neutral fraction of the foreground intergalactic medium \citep[IGM;][]{gunn65, fan06araa}. 
The luminosity and mass functions of quasars reflect the seeding and growth mechanisms of supermassive black holes (SMBHs), which can be studied through comparison with theoretical models \citep[e.g.,][]{volonteri12, ferrara14, madau14}. 
Measurements of quasar host galaxies and surrounding environments tell us about the earliest mass assembly of galaxies, which may happen in the highest-density peaks of the underlying dark matter distribution \citep[e.g.,][]{goto09, decarli17}.


For the past several years, we have been carrying out a project to search for high-$z$ quasars, named ``Subaru High-$z$ Exploration of Low-Luminosity 
Quasars (SHELLQs)'', with the Subaru 8.2-m telescope.
\citet{p1, p2, p4, p7} have already reported the discovery of 65 high-$z$ quasars, along with 24 high-$z$ luminous galaxies, 6 [\ion{O}{3}] emitters at $z \sim 0.8$, and 43 Galactic cool dwarfs (low-mass stars and brown dwarfs).
These objects include many low-luminosity quasars with redshifts up to $z = 7.07$, objects that have been difficult to find in past shallower surveys. 
Combined with samples of luminous quasars from the SDSS \citep{jiang16} and the Canada-France High-$z$ Quasar Survey \citep{willott10}, 
the SHELLQs sample has allowed us to establish the quasar luminosity function (LF) at $z = 6$ over an unprecedented range of magnitude \citep[$-22 < M_{1450} < -30$ mag;][]{p5}.
We found that the LF has a break at $M_{1450} \sim -25$ mag and flattens significantly toward the faint end.
With this LF shape, we predict that quasars are responsible for less than 10\% of 
the ionizing photons that are necessary to keep the IGM fully ionized at $z = 6$.

We are also carrying out multi-wavelength follow-up observations of the discovered quasars.
Near-infrared (IR) spectra were obtained for 6 quasars, which revealed that the SHELLQs quasars have 
a wide range of accretion properties, from sub-Eddington accretion onto a massive ($M_{\rm BH} > 10^9 M_\odot$) black hole to 
Eddington accretion onto a less massive ($M_{\rm BH} < 10^8 M_\odot$) black hole \citep{p6}.
Atacama Large Millimeter / submillimeter Array (ALMA) observations were conducted toward 7 quasars, and we found that the host galaxies are 
forming stars at a rate below that of luminous quasar hosts \citep[e.g.,][]{decarli18}, and that the low-luminosity quasars are more or less on the local 
relation between black hole mass and host mass \citep{p3, p8}.
We are also exploiting archival Wide-field Infrared Survey Explorer data to look for red quasars in the SHELLQs sample, which may 
represent a younger phase of evolution than normal quasars (N. Kato et al., in prep.).

This paper is the tenth in a series of SHELLQs publications, and reports spectroscopic identification of an additional 67 objects.
The paper is structured as follows.
We give brief descriptions of the data and methods used for candidate selection in \S \ref{sec:selection}. 
Spectroscopic follow-up observations are presented in \S \ref{sec:spectroscopy}.
Results and discussion
appear in \S \ref{sec:results}.
We adopt the cosmological parameters $H_0$ = 70 km s$^{-1}$ Mpc$^{-1}$, $\Omega_{\rm M}$ = 0.3, and $\Omega_{\rm \Lambda}$ = 0.7.
All magnitudes in the optical and near-IR bands are presented in the AB system \citep{oke83}, and are corrected for Galactic extinction \citep{schlegel98}.
We use the point spread function (PSF) magnitude ($m_{\rm AB}$) and the CModel magnitude ($m_{\rm CModel, AB}$), which are measured by fitting the PSF models and two-component, 
PSF-convolved galaxy models to the source profile, respectively \citep{abazajian04}.
In what follows, we refer to $z$-band magnitudes with the AB subscript (``$z_{\rm AB}$"), while redshift $z$ appears without a subscript.

\section{Photometric Candidate Selection} \label{sec:selection}

The SHELLQs project is based on the multi-band imaging data produced by the Hyper Suprime-Cam (HSC) Subaru Strategic 
Program (SSP) survey \citep{aihara17_survey}.
HSC is a wide-field camera on Subaru \citep{miyazaki18}.
It has a nearly circular field of view of 1$^\circ$.5 diameter,
covered by 116 2K $\times$ 4K Hamamatsu fully depleted CCDs.
The pixel scale is 0\arcsec.17.
The HSC-SSP survey has three layers with different combinations of area and depth.
The Wide layer is observing 1400 deg$^2$ mostly along the celestial equator, with 5$\sigma$ point-source depths of 
($g_{\rm AB}$, $r_{\rm AB}$, $i_{\rm AB}$, $z_{\rm AB}$, $y_{\rm AB}$) = (26.5, 26.1, 25.9, 25.1, 24.4) mag measured in 2\arcsec.0 aperture.
The total exposure times range from 10 minutes in the $g$- and $r$-bands to 20 minutes in the $i$-, $z$-, and $y$-bands, divided into individual exposures of $\sim$3 minutes each.
The Deep and the UltraDeep layers are observing smaller areas (27 and 3.5 deg$^2$) down to deeper limiting magnitudes ($r_{\rm AB}$ = 27.1 and 27.7 mag, respectively).
The survey data are reduced with the dedicated pipeline {\it hscPipe} \citep{bosch18} derived from the Large Synoptic Survey Telescope software pipeline \citep{juric17}.

The procedure of our quasar candidate selection has been described in the previous papers \citep{p1, p2, p4}, so here 
we repeat the essence only briefly.
The selection starts from the HSC-SSP survey data in the three layers, over the area observed in the $i$, $z$, and $y$ bands for $i$-dropout selection
or in the $z$ and $y$ bands for $z$-dropout selection.
We include data that haven't yet reached the planned full depth.
All sources meeting the following criteria, and that are not flagged as having suspect detection or photometry, are selected:
\begin{eqnarray}
(z_{\rm AB} < 24.5\ {\rm and}\ \sigma_z < 0.155\ {\rm and}\ i_{\rm AB} - z_{\rm AB} > 1.5 \nonumber \\
 {\rm and}\ z_{\rm AB} - z_{\rm CModel, AB} < 0.15 )
\label{eq:query1}
\end{eqnarray}
or
\begin{eqnarray}
(y_{\rm AB} < 25.0\ {\rm and}\ \sigma_y < 0.217\ {\rm and}\ z_{\rm AB} - y_{\rm AB} > 0.8 \nonumber \\
 {\rm and}\ y_{\rm AB} - y_{\rm CModel, AB} < 0.15 ) .
\label{eq:query2}
\end{eqnarray}
Equations (\ref{eq:query1}) and (\ref{eq:query2}) select $i$-dropout and $z$-dropout candidates, respectively.
Here ($\sigma_i$, $\sigma_z$, $\sigma_y$) refer to the errors of the PSF magnitudes ($i_{\rm AB}$, $z_{\rm AB}$, $y_{\rm AB}$), as measured by {\it hscPipe}.
The $y$-band limiting magnitude in Equation (\ref{eq:query2}) is deeper than the limit used in our previous papers ($y_{\rm AB} < 24.0$), as we are shifting our focus
to establishing the quasar LF at $z \sim 7$.
The difference between the PSF and CModel magnitudes ($m_{\rm AB} - m_{\rm CModel, AB}$) is used to exclude extended sources.
Sources with more than $3\sigma$ detection in the $g$ or $r$ band (if these bands are available) are removed as likely low-$z$ interlopers.
We also remove those sources whose images appear to be moving objects, transients, or artifacts, 
spotted by automatic or eye inspection \citep[see][]{p1}.
We always keep ambiguous cases in our sample for spectroscopic follow-up, so that we do not discard any real quasars in the selection.


The candidates selected from the HSC data are matched, within 1\arcsec.0 separation, to the public near-IR catalogs from
the United Kingdom Infrared Telescope Infrared Deep Sky Survey \citep[UKIDSS;][]{lawrence07} data release (DR) 10 and 11,  
the Visible and Infrared Survey Telescope for Astronomy (VISTA) Kilo-degree Infrared Galaxy survey (VIKING) DR 4 and 5, and
the VISTA Deep Extragalactic Observations Survey \citep{jarvis13} DR 5.
In practice, only a small fraction of the final quasar candidates have near-IR photometry, due to limited sky coverage and/or depth of the above surveys.
Using all available magnitudes in the $i$, $z$, $y$, $J$, $H$, and $K$ bands, we calculate the Bayesian quasar 
probability ($P_{\rm Q}^{\rm B}$) for each candidate.
The calculation is based on models of spectral energy distributions and surface densities of high-$z$ quasars and 
contaminating brown dwarfs, as a function of magnitude; 
galaxy models are not included in the algorithm at present \citep[see][]{p1}.
We keep those sources with $P_{\rm Q}^{\rm B} > 0.1$ in the sample of candidates, while removing sources with lower quasar probabilities.

We used the latest HSC-SSP (internal) DR, which includes observations carried out between 2014 March and 2018 January. 
The data cover roughly 900 deg$^2$, when we limit to the fields where at least (1, 2, 2) exposures were taken in the ($i$, $z$, $y$) bands 
for $i$-dropout selection, or at least (1, 2) exposures were taken in the ($z$, $y$) bands for $z$-dropout selection.
We found $\sim$500 final candidates over this area, and put highest priority on (i) all the $z$-dropouts and (ii) the $i$-band dropouts with $i_{\rm AB} - z_{\rm AB} > 2.0$,
$z_{\rm AB} <$ 24.0 mag, and $y$-band detection.
We have almost completed follow-up spectroscopy of these $\sim$200 high-priority candidates, as reported in the past and present papers, and continue to observe the remaining
low-priority candidates.

\section{Spectroscopy} \label{sec:spectroscopy}

We carried out follow-up spectroscopy of 67 candidates from 2018 March through 2019 July, 
using the Optical System for Imaging and low-intermediate-Resolution Integrated Spectroscopy \citep[OSIRIS;][]{cepa00} mounted on the 10.4-m Gran Telescopio Canarias (GTC),
and the Faint Object Camera and Spectrograph \citep[FOCAS;][]{kashikawa02} mounted on Subaru.
We prioritized observations in such a way that the targets with brighter magnitudes and higher $P_{\rm Q}^{\rm B}$ were observed at earlier opportunities.
Table \ref{tab:obsjournal} is a journal of the spectroscopic observations.


GTC is a 10.4-m telescope located at the Observatorio del Roque de los Muchachos in La Palma, Spain.
We used OSIRIS with the R2500I grism and 1\arcsec.0-wide longslit, which provides spectral coverage from $\lambda_{\rm obs}$ = 0.74 to 1.0\ $\mu$m 
with a resolution $R \sim 1500$.
The observations (Program IDs: GTC3-18A, GTC8-18B, and GTC32-19A) were carried out in queue mode on dark and gray nights, under spectroscopic sky conditions and 
with seeing 0\arcsec.6 -- 1\arcsec.3.

The Subaru observations were carried out as part of two Subaru Intensive Programs (Program IDs: S16B-071I and S18B-011I).
We used FOCAS in the multi-object spectrograph (MOS) mode with the VPH900 grism and SO58 order-sorting filter.
The widths of the slitlets were set to 1\arcsec.0.
This configuration provides spectral coverage from $\lambda_{\rm obs}$ = 0.75 to 1.05\ $\mu$m with a resolution $R \sim 1200$.
All the observations were carried out on gray nights, which were occasionally affected by cirrus,
with seeing 0\arcsec.4 -- 1\arcsec.0.

All the data obtained with GTC and Subaru were reduced using the Image Reduction and Analysis Facility (IRAF).
Bias correction, flat fielding with dome flats, sky subtraction, and 1d extraction were performed in the standard way.
The wavelength was calibrated with reference to sky emission lines.
The flux calibration was tied to white dwarfs (Feige 34, Feige 110, G191-B2B, GD 153, Ross 640) or a B-type standard star (HILT 600), observed 
as standard stars within a few days of the target observations.
We corrected for slit losses by scaling the spectra to match the HSC magnitudes in the $z$ and $y$ bands for the $i$- and $z$-band dropouts, respectively.

\startlongtable
\begin{deluxetable*}{lcccrrl}
\tablecaption{Journal of Discovery Spectroscopy \label{tab:obsjournal}}
\tablehead{
\colhead{Object} & 
\colhead{$i_{\rm AB}$} & \colhead{$z_{\rm AB}$} & \colhead{$y_{\rm AB}$} & \colhead{$P_{\rm Q}^{\rm B}$} & 
\colhead{$t_{\rm exp}$} & \colhead{Date (Inst)}\\
\colhead{} & 
\colhead{(mag)} & \colhead{(mag)} & \colhead{(mag)} & \colhead{} &
\colhead{(min)} & \colhead{}
} 
\startdata
\multicolumn{7}{c}{Quasars}\\\hline
$J235646.33+001747.3$ &  $>$26.51              & 27.20 $\pm$ 1.07 & 22.94 $\pm$ 0.05 & 1.000 & 120 & 2018 Nov 7, 8 (O)\\
$J160953.03+532821.0$ &  $>$26.89              &          $>$26.317   & 24.15 $\pm$ 0.09 & 1.000  & 180 &  2019 Jul 4, 24 (O)   \\ 
$J011257.84+011042.4$ &  $>$25.20              & $>$24.27              & 23.04 $\pm$ 0.06 & 0.222 & 180 & 2018 Nov 8, 30 (O)\\
$J161207.12+555919.2$ &          $>$27.21      &          $>$26.38     & 24.09 $\pm$ 0.07 &  1.000 & 180 & 2019 Jun 25 (O)     \\ 
$J134400.87+012827.8$ &  26.85 $\pm$ 0.44 & $>$25.25              & 22.89 $\pm$ 0.07 & 1.000 & 30 & 2018 Apr 24 (F)\\
$J000142.54+000057.5$ &  27.21 $\pm$ 0.49 & 26.56 $\pm$ 0.73 & 22.59 $\pm$ 0.03 & 1.000 & 45 & 2018 Oct 13 (O)\\
$J123137.77+005230.3$ &  27.29 $\pm$ 0.42 & 25.96 $\pm$ 0.26 & 22.76 $\pm$ 0.05 & 1.000 & 30 & 2018 Apr 24 (F)\\
$J135012.04-002705.2$ &  26.71 $\pm$ 0.29 & 23.33 $\pm$ 0.03 & 22.89 $\pm$ 0.05 & 1.000 & 60 & 2018 Mar 17 (O)\\
$J084456.62+022640.5$ & 27.75 $\pm$ 0.98 & 23.83 $\pm$ 0.06 & 25.42 $\pm$ 0.33 & 1.000  & 30 & 2019 May 11 (F) \\ 
$J113753.64+004509.7$ & 26.89 $\pm$ 0.42 & 23.42 $\pm$ 0.03 & 22.30 $\pm$ 0.03 & 0.899 & 60 & 2018 Mar 11 (O)\\
$J152555.79+430324.0$ & 26.53 $\pm$ 0.20 & 23.34 $\pm$ 0.03 & 23.48 $\pm$ 0.09 & 1.000 & 60 & 2018 Mar 13 (O)\\
$J151248.71+442217.5$ &  26.15 $\pm$ 0.23 & 24.63 $\pm$ 0.14 & 24.16 $\pm$ 0.17 & 0.000 & 30 & 2018 Apr 24 (F)\\
$J225520.78+050343.3$ &  26.26 $\pm$ 0.39 & 22.31 $\pm$ 0.02 & 22.40 $\pm$ 0.04 & 1.000 & 30 & 2018 Aug 3 (O)\\
$J134733.69-015750.6$ &  25.12 $\pm$ 0.10 & 21.65 $\pm$ 0.01 & 22.12 $\pm$ 0.03 & 1.000 & 15 & 2019 Apr 24 (F)\\
$J144823.33+433305.9$ & $>$24.22              & 22.33 $\pm$ 0.03 & 22.46 $\pm$ 0.04 & 1.000 & 30 & 2018 Aug 11 (O)\\
$J000133.30+000605.4$ & 26.94 $\pm$ 0.34 & 23.68 $\pm$ 0.05 & 23.57 $\pm$ 0.07 & 1.000 & 120 &  2019 Jul 1, 4 (O)    \\ 
$J151657.87+422852.9$ & 25.85 $\pm$ 0.16 & 22.47 $\pm$ 0.01 & 22.46 $\pm$ 0.03 & 1.000 & 50 & 2018 Apr 25 (F)\\
$J125437.08-001410.7$ &          $>$25.45   & 23.89 $\pm$ 0.07 & 25.45 $\pm$ 0.75 &  1.000  & 30 & 2019 May 11 (F) \\ 
$J000445.81-004944.3$ &  25.53 $\pm$ 0.12 & 22.54 $\pm$ 0.02 & 22.81 $\pm$ 0.05 & 1.000 & 30 & 2018 Aug 3 (O)\\
$J093543.32-011033.3$ & 26.79 $\pm$ 0.33 & 23.50 $\pm$ 0.04 & 24.79 $\pm$ 0.30 & 1.000 & 30 & 2019 May 10 (F)   \\ 
$J010603.68-003015.2$ &  25.89 $\pm$ 0.43 & 22.83 $\pm$ 0.04 & 23.14 $\pm$ 0.06 & 1.000 & 45 & 2018 Aug 9 (O)\\
$J142611.33-012822.8$ &  27.38 $\pm$ 0.75 & 23.70 $\pm$ 0.06 & 23.56 $\pm$ 0.15 & 1.000 & 120 & 2019 Apr 9 (O)\\
$J121049.13+013426.7$ & 25.95 $\pm$ 0.26 & 23.90 $\pm$ 0.07 & 23.95 $\pm$ 0.24 & 0.900 & 30 & 2019 May 10 (F) \\ 
$J093000.85+005738.4$ & 23.51 $\pm$ 0.02 & 21.72 $\pm$ 0.01 & 21.62 $\pm$ 0.02 & 1.000 & 15 & 2019 Mar 8 (O)\\
$J161819.77+552654.0$ & 24.14 $\pm$ 0.02 & 22.43 $\pm$ 0.01 & 22.41 $\pm$ 0.02 & 1.000 & 40 & 2018 Apr 25 (F)\\
$J142307.21-022519.0$ &  24.28 $\pm$ 0.10 & 22.49 $\pm$ 0.05 & 22.85 $\pm$ 0.22 & 1.000 & 15 & 2018 Aug 11 (O)\\
$J023858.09-031845.4$ &  24.17 $\pm$ 0.08 & 22.64 $\pm$ 0.04 & 22.53 $\pm$ 0.08 & 1.000 & 30 & 2018 Sep 3 (O)\\
$J113218.15+003800.1$ & 25.90 $\pm$ 0.29 & 23.66 $\pm$ 0.08 & 23.73 $\pm$ 0.11 & 1.000 & 100 & 2019 Apr 24 (F)\\
\hline\multicolumn{7}{c}{Galaxies}\\\hline
$J135348.55-001026.5$ &  27.25 $\pm$ 0.44 & 23.33 $\pm$ 0.03 & 22.65 $\pm$ 0.03 & 1.000 & 60 & 2018 Mar 11 (O)\\
$J144216.08+423632.5$ & 26.14 $\pm$ 0.30 & 23.91 $\pm$ 0.07 & 23.63 $\pm$ 0.11 & 0.232 & 30 & 2019 May 10 (F) \\ 
$J092117.65+030521.5$ & 26.69 $\pm$ 0.26 & 24.05 $\pm$ 0.06 & 23.84 $\pm$ 0.14 & 1.000 & 30 & 2019 May 11 (F)   \\ 
$J115755.51-001356.2$ &  26.53 $\pm$ 0.23 & 23.91 $\pm$ 0.05 & 23.80 $\pm$ 0.09 & 1.000 & 40 & 2019 Apr 26 (F)\\
$J123841.97-011738.8$ &  24.99 $\pm$ 0.14 & 23.46 $\pm$ 0.03 & 23.56 $\pm$ 0.08 & 1.000 & 30 & 2019 Apr 24 (F)\\
$J162657.22+431133.0$ & 26.24 $\pm$ 0.14 & 24.02 $\pm$ 0.07 & 24.30 $\pm$ 0.16 & 1.000 & 50 & 2019 May 10 (F)   \\ 
$J020649.98-020618.2$ &  25.05 $\pm$ 0.13 & 23.20 $\pm$ 0.03 & 23.11 $\pm$ 0.06 & 1.000 & 60 & 2018 Sep 8 (O)\\
\hline\multicolumn{7}{c}{[\ion{O}{3}] Emitters}\\\hline
$J090017.67-014655.6$ & 24.93 $\pm$ 0.09 & 23.30 $\pm$ 0.04 & 23.61 $\pm$ 0.14 & 1.000 & 60 & 2019 Mar 13 (O)\\
$J015519.63-005814.7$ & $>$22.80              & 23.50 $\pm$ 0.07 & 23.56 $\pm$ 0.08 & 0.907 & 120 & 2018 Nov 3 (O)\\
$J144221.95-013258.0$ & 26.06 $\pm$ 0.20 & 23.98 $\pm$ 0.07 & 24.86 $\pm$ 0.25 & 1.000 & 10 & 2019 May 10 (F)  \\ 
$J115946.07+014023.7$ & 22.69 $\pm$ 0.01 & 21.14 $\pm$ 0.01 & 22.94 $\pm$ 0.06 & 1.000 & 15 & 2019 Jun 9 (O)    \\ 
$J141332.93+014350.4$ & 25.79 $\pm$ 0.17 & 23.65 $\pm$ 0.08 &          $>$24.39 & 1.000 & 10 & 2019 May 10 (F)   \\ 
$J094209.69-020302.1$ & 23.57 $\pm$ 0.02 & 21.91 $\pm$ 0.02 & 23.34 $\pm$ 0.12 & 1.000 & 15 & 2019 Mar 8 (O)\\
$J225124.84-010911.0$ & 24.32 $\pm$ 0.05 & 22.69 $\pm$ 0.03 & 24.53 $\pm$ 0.19 & 1.000 & 42 & 2018 Sep 1 (O)\\
$J093831.54-002523.4$ & 25.98 $\pm$ 0.15 & 23.67 $\pm$ 0.05 & 26.12 $\pm$ 1.00 & 1.000 & 10 & 2019 May 10 (F)   \\ 
$J135049.56+013726.6$ & 25.94 $\pm$ 0.21 & 23.94 $\pm$ 0.06 &          $>$24.08 & 1.000 & 10 & 2019 May 10 (F)   \\ 
$J150602.11+415317.5$ & 23.54 $\pm$ 0.03 & 21.82 $\pm$ 0.01 & 24.11 $\pm$ 0.22 & 1.000 & 20 & 2018 Sep 1 (O)\\
\hline\multicolumn{7}{c}{Cool Dwarfs}\\\hline
$J011152.80+013211.4$ & 25.86 $\pm$ 0.43 & 23.54 $\pm$ 0.14 & 23.67 $\pm$ 0.08 & 0.218 & 120 & 2018 Oct 13, 15 (O)\\
$J015548.62-061737.9$ & 24.66 $\pm$ 0.16 & 23.12 $\pm$ 0.05 & 23.01 $\pm$ 0.10 & 0.242 & 60 & 2018 Sep 8 (O)\\
$J020421.81-024421.1$ & 23.07 $\pm$ 0.02 & 20.21 $\pm$ 0.01 & 19.19 $\pm$ 0.01 & 1.000  & 20 & 2018 Sep 1 (O)\\
$J022144.40-053008.6$ & 25.94 $\pm$ 0.15 & 22.93 $\pm$ 0.02 & 21.86 $\pm$ 0.02 & 0.410 & 60 & 2018 Nov 9 (O)\\
$J085230.35+025410.0$ & $>$25.31              & 24.00 $\pm$ 0.07 & 23.86 $\pm$ 0.13 & 0.996 & 50 & 2019 Apr 25 (F)\\
$J085818.99+040927.6$ & $>$25.50              & 23.54 $\pm$ 0.05 & 22.85 $\pm$ 0.08 & 0.929 & 120 & 2018 Nov 3 (O)\\
$J124505.57+010550.2$ & \nodata                 & 22.05 $\pm$ 0.02 & 21.20 $\pm$ 0.01 & \nodata & 20 & 2018 Apr 25 (F)\\
$J135913.21+002740.0$ & 26.75 $\pm$ 0.30 & 24.04 $\pm$ 0.06 & 22.16 $\pm$ 0.02 & 0.000 & 40 & 2019 Apr 25 (F)\\
$J142531.82-021606.5$ &  26.68 $\pm$ 0.62 & 23.50 $\pm$ 0.08 & 22.90 $\pm$ 0.17 & 0.000 & 120 & 2019 Apr 9 (O)\\
$J145853.06+015031.8$ & 24.05 $\pm$ 0.03 & 21.42 $\pm$ 0.02 & 20.40 $\pm$ 0.01 & 0.000 & 15 & 2018 Aug 10 (O)\\
$J151812.04+440829.0$ & 26.30 $\pm$ 0.20 & 23.38 $\pm$ 0.02 & 22.30 $\pm$ 0.02 & 0.062 & 60 & 2018 Mar 11 (O)\\
$J153032.76+435615.5$ & 25.27 $\pm$ 0.07 & 22.49 $\pm$ 0.01 & 21.55 $\pm$ 0.01 & 0.998 & 30 & 2018 Aug 11 (O)\\
$J154605.67+425306.1$ & 28.07 $\pm$ 0.83 & 24.94 $\pm$ 0.13 & 22.64 $\pm$ 0.05 & 0.790 & 100 & 2019 Apr 24 (F)\\
$J161042.47+554203.4$ & 26.88 $\pm$ 0.17 & 24.68 $\pm$ 0.05 & 23.61 $\pm$ 0.05 & 1.000 & 141 & 2019 Jun 26 (O)    \\ 
$J164226.30+430749.7$ & 25.00 $\pm$ 0.12 & 23.35 $\pm$ 0.05 & 23.48 $\pm$ 0.20 & 0.936 & 66 & 2019 Mar 8 (O)\\
$J223827.66+053246.5$ & 24.82 $\pm$ 0.13 & 23.13 $\pm$ 0.04 & 23.08 $\pm$ 0.07 & 1.000 & 60 & 2018 Sep 2 (O)\\
$J224157.52+055912.6$ & 24.52 $\pm$ 0.09 & 22.88 $\pm$ 0.04 & 23.06 $\pm$ 0.08 & 1.000 & 30 & 2018 Aug 3 (O)\\
$J225132.39-011601.1$ &  24.46 $\pm$ 0.08 & 22.75 $\pm$ 0.07 & 22.91 $\pm$ 0.06 & 0.994 & 30 & 2018 Aug 3 (O)\\
$J225337.55+051440.1$ & $>$25.32              & 23.52 $\pm$ 0.05 & 22.77 $\pm$ 0.06 & 0.152 & 60 & 2018 Sep 5 (O)\\
$J225513.58+044226.8$ &    \nodata              & 22.59 $\pm$ 0.03 & 21.92 $\pm$ 0.05 & \nodata & 30 & 2018 Aug 3 (O)\\
$J230811.42+023931.3$ & 24.12 $\pm$ 0.05 & 22.40 $\pm$ 0.02 & 22.08 $\pm$ 0.06 & 0.998 & 15 & 2018 Aug 3 (O)\\
$J231344.68-003931.3$ &  24.85 $\pm$ 0.06 & 23.17 $\pm$ 0.03 & 22.94 $\pm$ 0.07 & 1.000 & 60 & 2018 Sep 1 (O)\\
\enddata
\tablecomments{
Coordinates are at J2000.0, and magnitude upper limits are placed at $5\sigma$ significance. We took magnitudes from the latest HSC-SSP DR, 
and recalculated $P_{\rm Q}^{\rm B}$ for objects selected from older DRs; this is why a few objects have $P_{\rm Q}^{\rm B} < 0.1$.
No $i$-band images are available at the positions of $J124505.57+010550.2$ and $J225513.58+044226.8$ in the latest DR, due to quality issues, 
and thus $i_{\rm AB}$ and $P_{\rm Q}^{\rm B}$ are not available for these sources.
The instrument (Inst) ``O" and ``F" denote GTC/OSIRIS and Subaru/FOCAS, respectively.}
\end{deluxetable*}

\begin{deluxetable*}{ccccl}
\tablecaption{Objects detected in the near-IR bands \label{tab:nir_photometry}}
\tablehead{
\colhead{Name} & \colhead{$J_{\rm AB}$} & \colhead{$H_{\rm AB}$} & \colhead{$K_{\rm AB}$} & \colhead{Survey}\\
\colhead{} & \colhead{(mag)} & \colhead{(mag)} & \colhead{(mag)} & \colhead{} 
} 
\startdata
\multicolumn{5}{c}{Quasars}\\\hline
$J113753.64+004509.7$ & 21.51 $\pm$ 0.20 & 21.67 $\pm$ 0.36 & \nodata & VIKING \\
$J161819.77+552654.0$ & $22.45 \pm 0.10$ & \nodata & $21.82 \pm 0.09$ &  UKIDSS (Deep Extragalactic Survey)\\
$J142307.21-022519.0$ & 22.17 $\pm$ 0.27 & \nodata & \nodata & VIKING \\
\hline\multicolumn{5}{c}{Galaxies}\\\hline
$J135348.55-001026.5$ & 22.15 $\pm$ 0.26 & \nodata & 20.78 $\pm$ 0.16 & VIKING \\
\hline\multicolumn{5}{c}{Cool Dwarfs}\\\hline
$J142531.82-021606.5$ & 21.54 $\pm$ 0.15 & 20.84 $\pm$ 0.16 & 20.46 $\pm$ 0.12 & VIKING \\
$J161042.47+554203.4$ & 23.14 $\pm$ 0.19 & \nodata & 23.25 $\pm$ 0.32 & UKIDSS (Deep Extragalactic Survey)\\
$J230811.42+023931.3$ & \nodata & 20.74 $\pm$ 0.26 & \nodata & UKIDSS (Large Area Survey) \\
\enddata
\end{deluxetable*}


\section{Results and Discussion \label{sec:results}}

\begin{figure*}
\epsscale{1.0}
\plotone{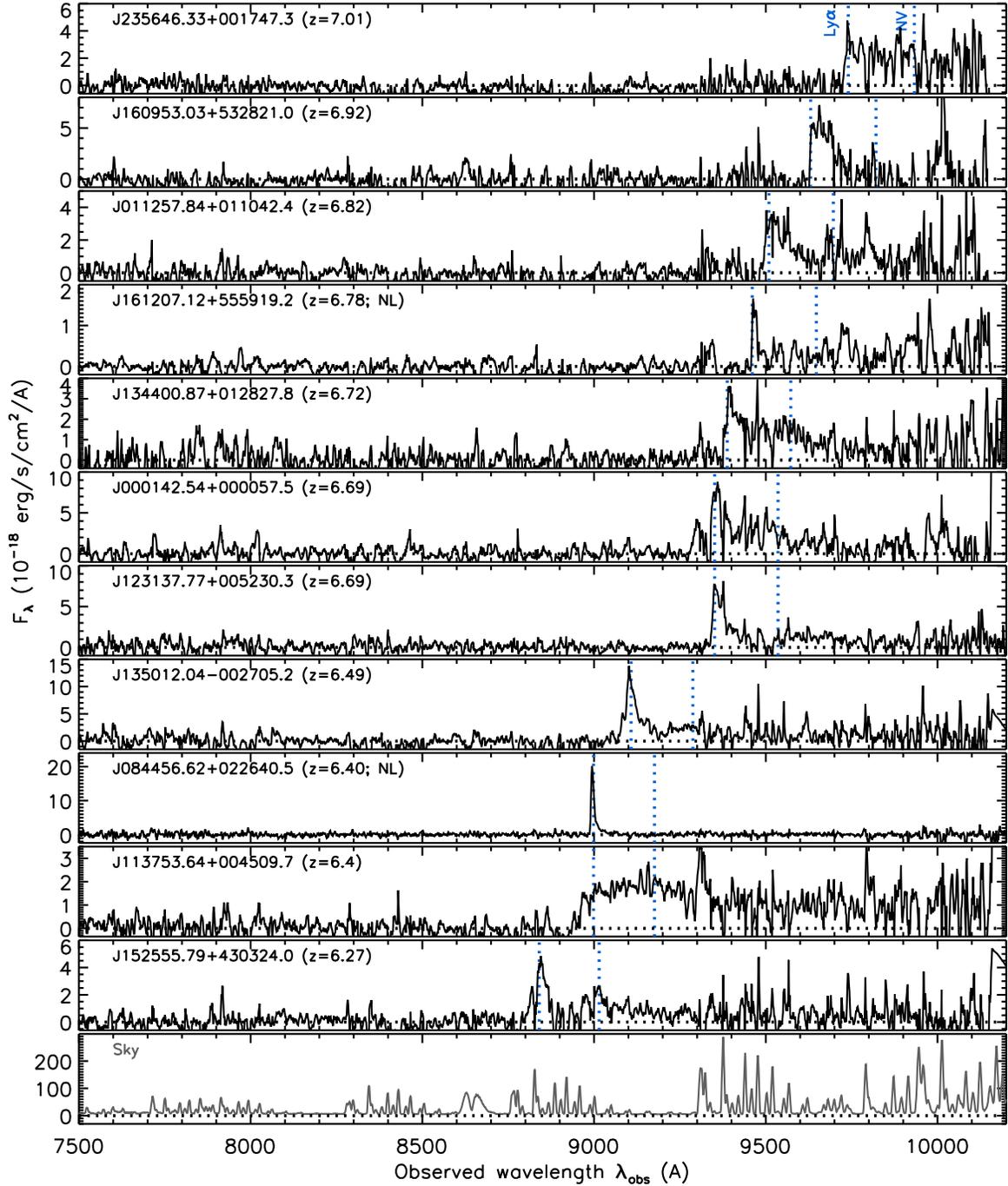}
\caption{Discovery spectra of the first set of 11 quasars, displayed in decreasing order of redshift.
The object name and the estimated redshift (and the designation ``NL" for the possible quasars with narrow Ly$\alpha$ emission) are indicated at the top left corner of each panel.
The blue dotted lines mark the expected positions of the Ly$\alpha$ and \ion{N}{5} $\lambda$1240 emission lines, given the redshifts.
The spectra were smoothed using inverse-variance weighted means over 1 -- 9 pixels (depending on the signal-to-noise ratio [S/N]), for display purposes.
The bottom panel displays a sky spectrum, as a guide to the expected noise.
\label{fig:spectra1}}
\end{figure*}

\begin{figure*}
\epsscale{1.0}
\plotone{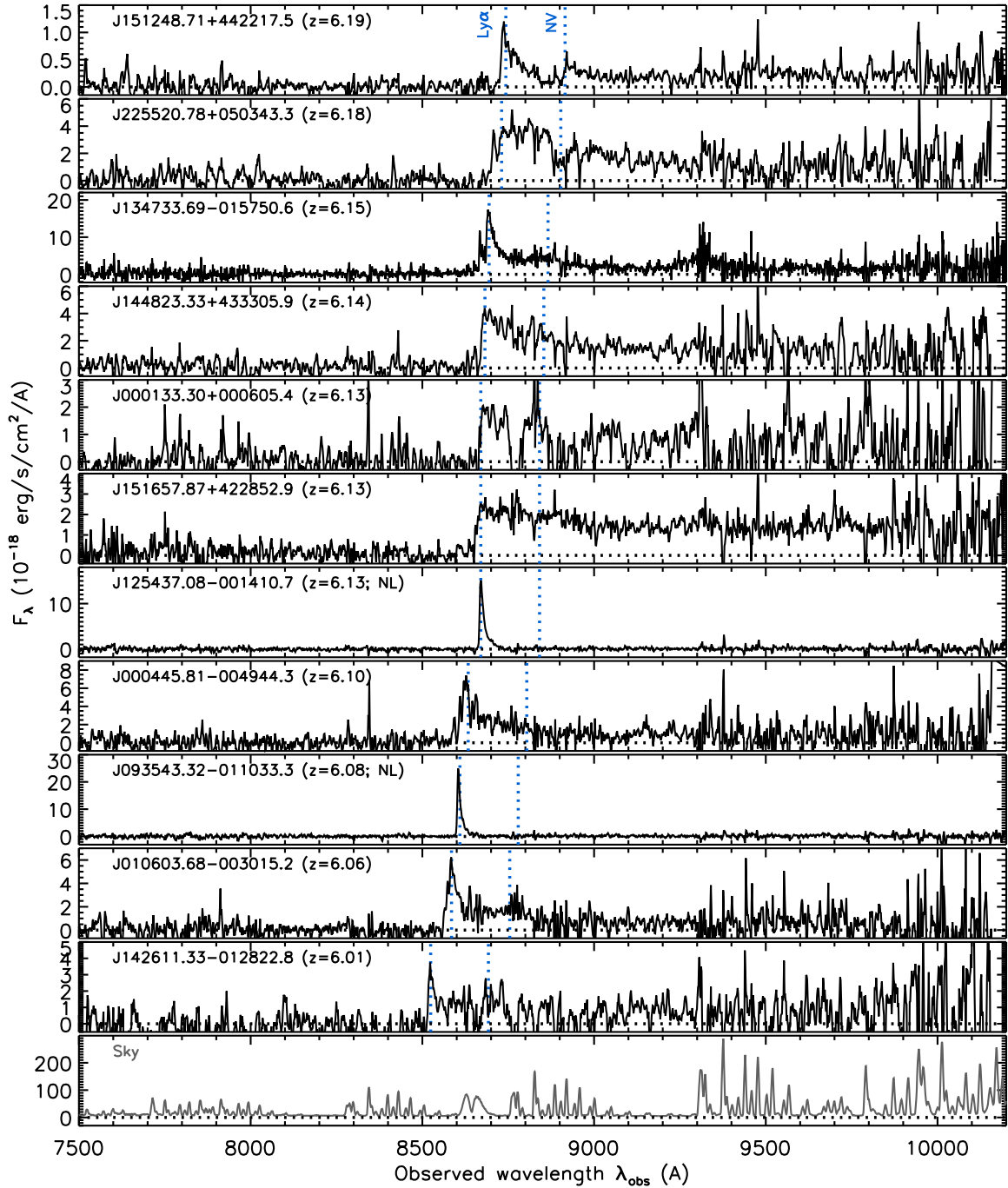}
\caption{Same as Figure \ref{fig:spectra1}, but for the second set of 11 quasars.
\label{fig:spectra2}}
\end{figure*}

\begin{figure*}
\epsscale{1.0}
\plotone{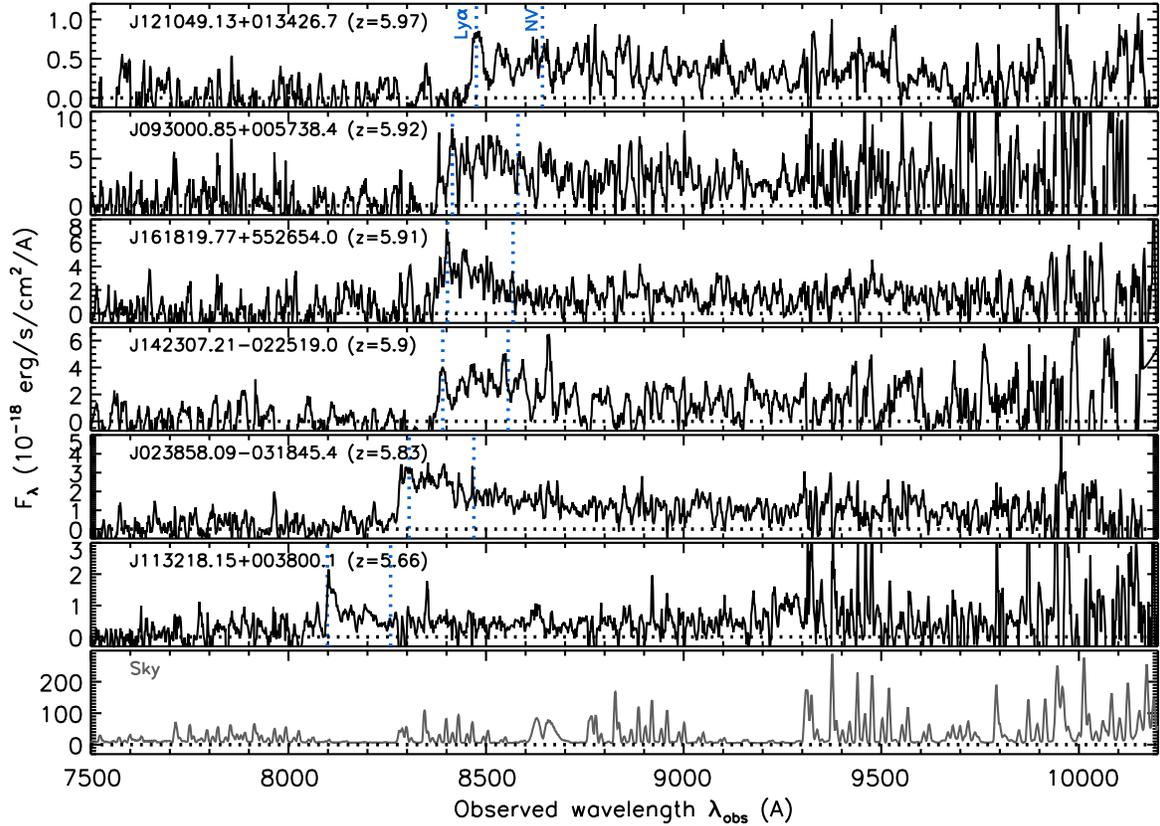}
\caption{Same as Figure \ref{fig:spectra1}, but for the last set of 6 quasars.
\label{fig:spectra3}}
\end{figure*}

\begin{figure*}
\epsscale{1.0}
\plotone{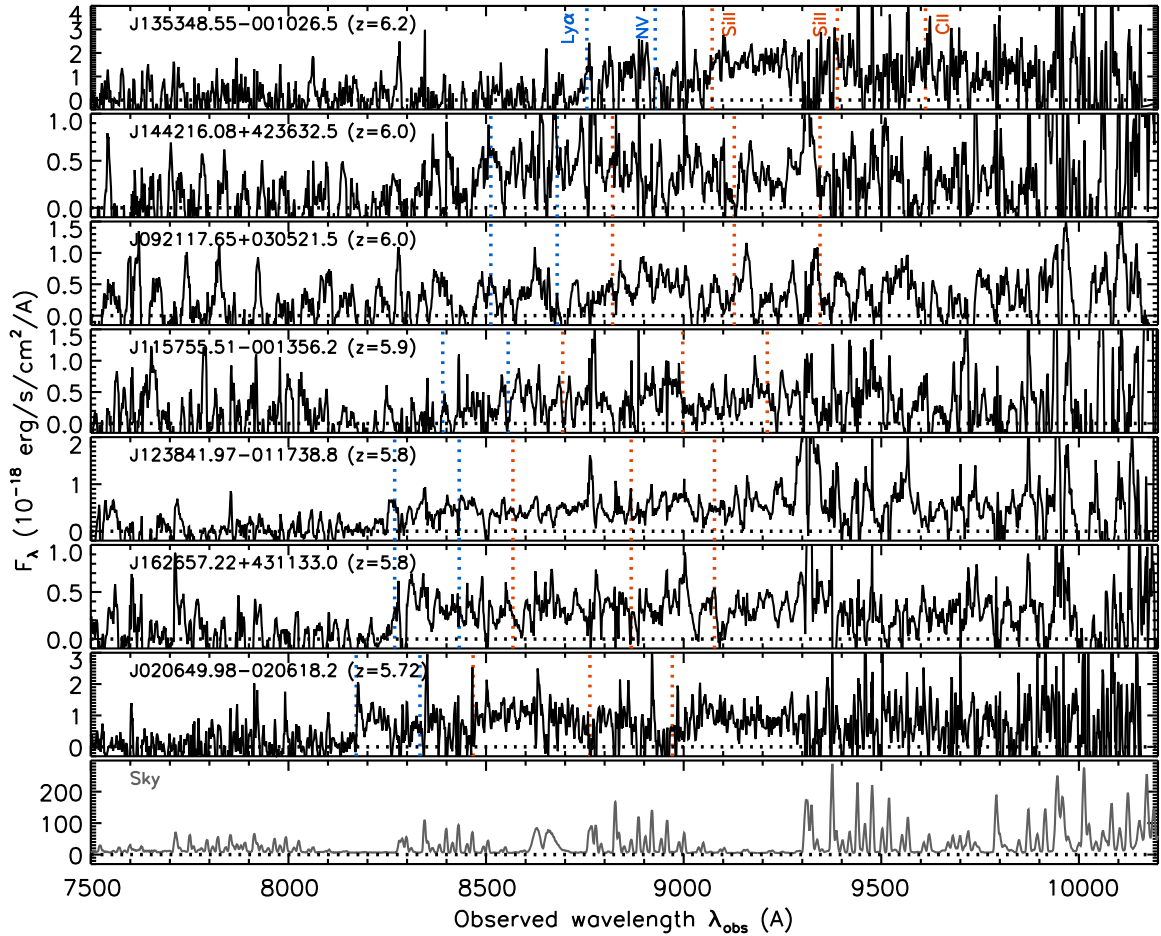}
\caption{Same as Figure \ref{fig:spectra1}, but for the 7 high-$z$ galaxies.
The expected positions of the interstellar absorption lines of \ion{Si}{2} $\lambda$1260, \ion{Si}{2} $\lambda$1304, and \ion{C}{2} $\lambda$1335 are marked by the red dotted lines.
\label{fig:spectra4}}
\end{figure*}

\begin{figure*}
\epsscale{1.0}
\plotone{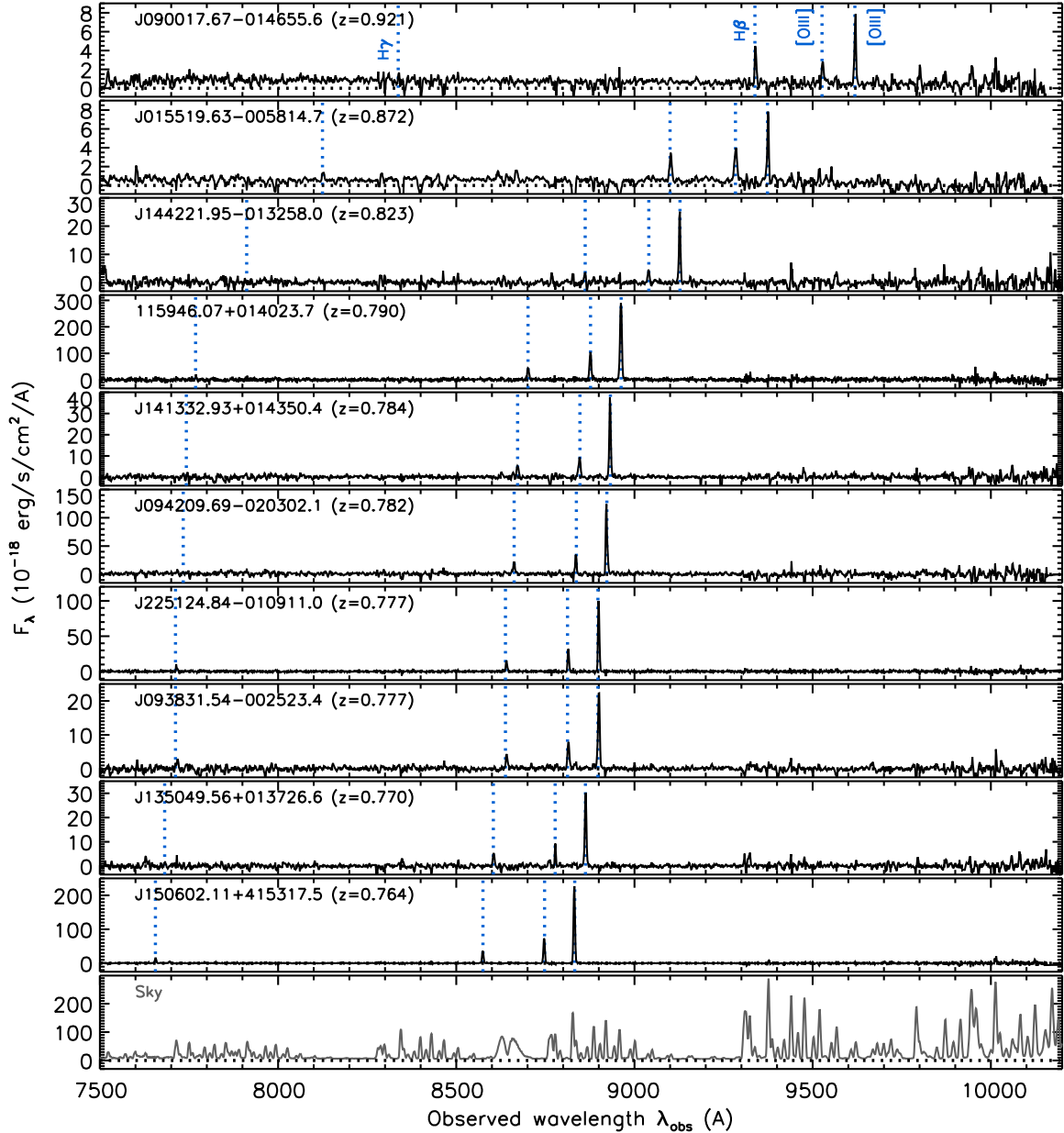}
\caption{Same as Figure \ref{fig:spectra1}, but for the 10 [\ion{O}{3}] emitters at $z \sim 0.8$.
The expected positions of H$\gamma$, H$\beta$, and two [\ion{O}{3}] lines ($\lambda$4959 and $\lambda$5007) are marked by the dotted lines.
\label{fig:spectra5}}
\end{figure*}

\begin{figure*}
\epsscale{1.0}
\plotone{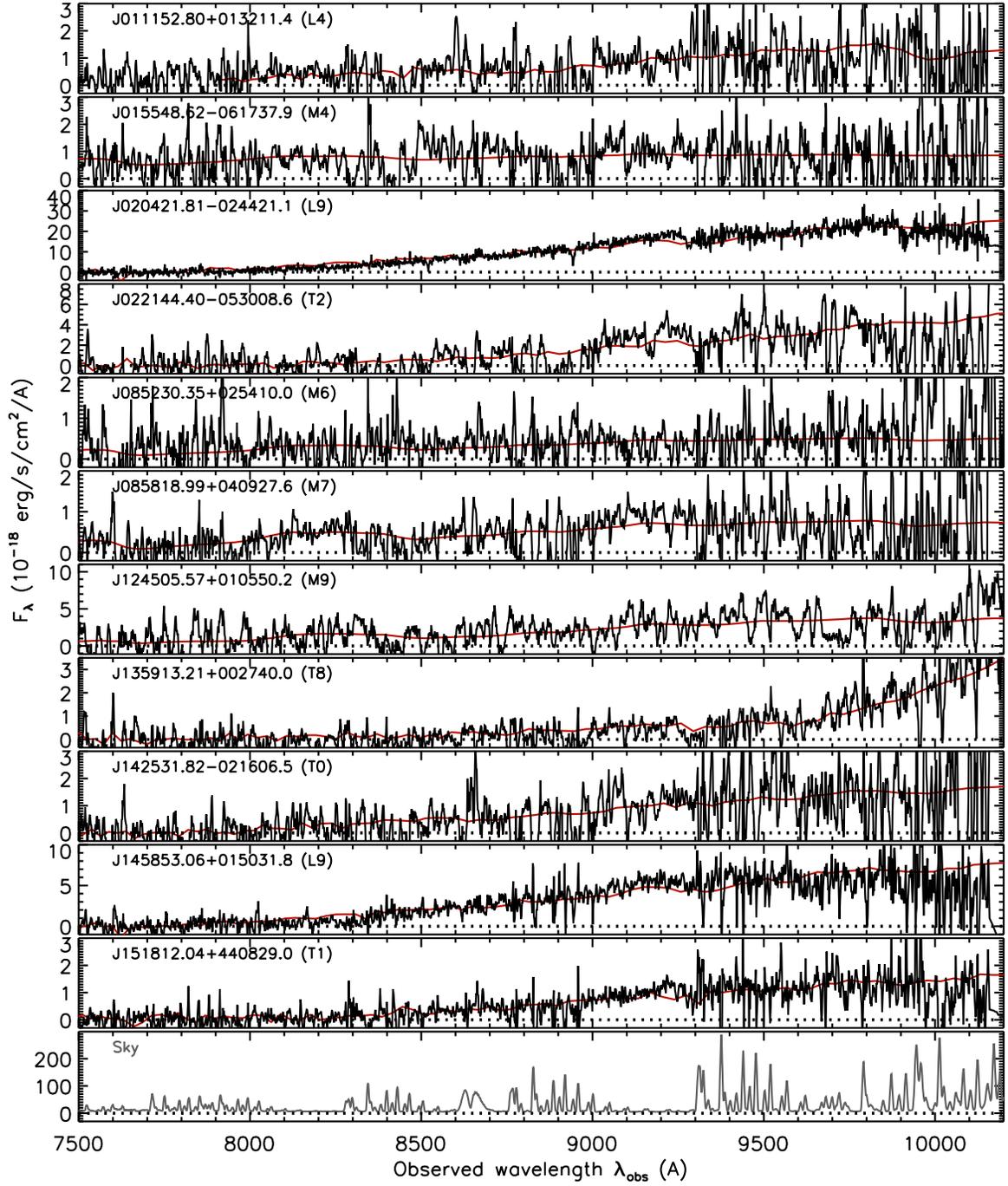}
\caption{Same as Figure \ref{fig:spectra1}, but for the first set of 11 cool dwarfs. 
The red lines represent the best-fit templates, whose spectral types are indicated at the top left corner of each panel.
The small-scale ($<$100 \AA) features seen in the spectra are due to noise in most cases, given the limited S/N.
\label{fig:spectra6}}
\end{figure*}

\begin{figure*}
\epsscale{1.0}
\plotone{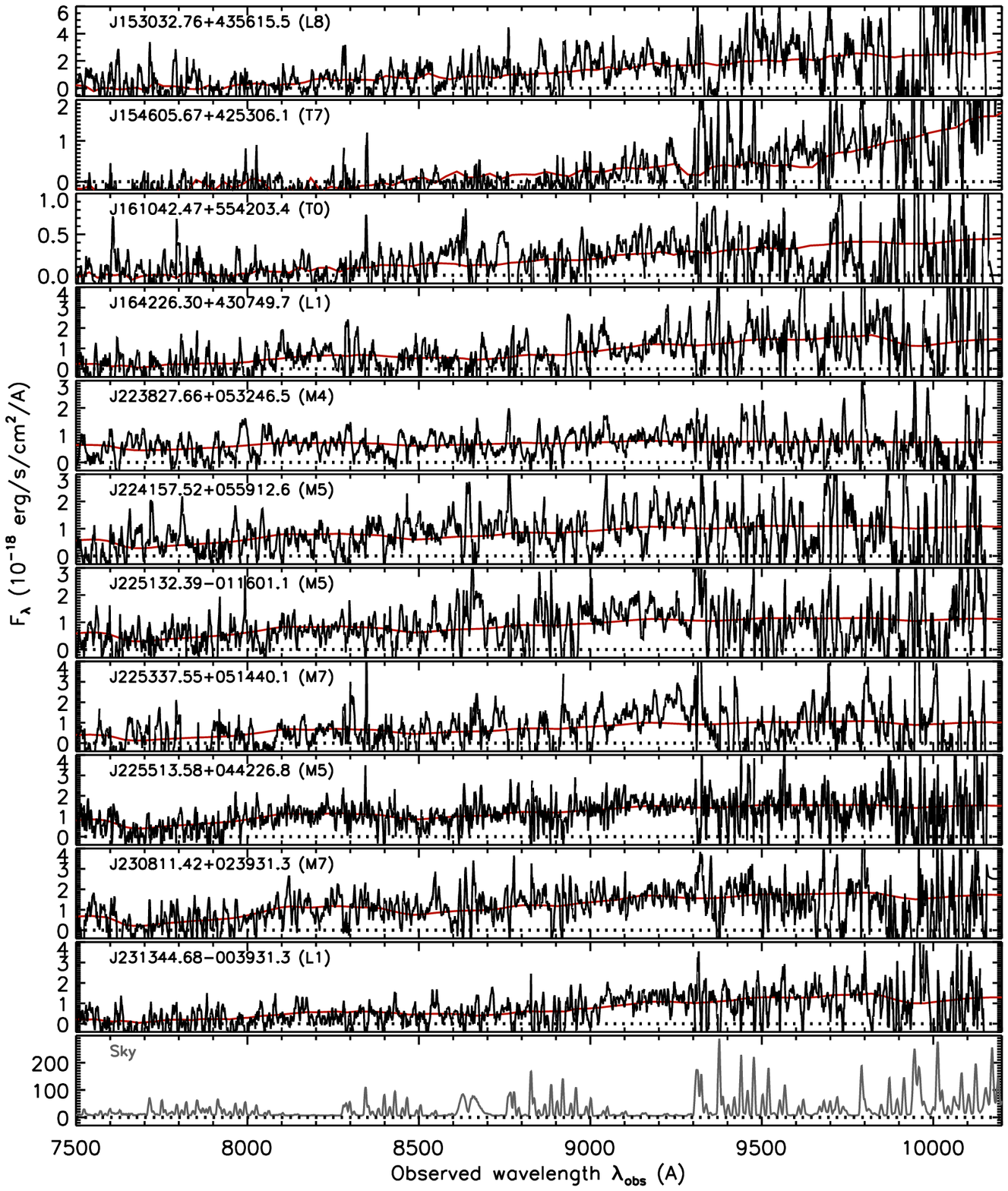}
\caption{Same as Figure \ref{fig:spectra6}, but for the second set of 11 cool dwarfs. 
\label{fig:spectra7}}
\end{figure*}

We present the reduced spectra in Figures \ref{fig:spectra1} -- \ref{fig:spectra7}.
Based on these spectra, we identified 28 high-$z$ quasars, 7 high-$z$ galaxies, 10 strong [\ion{O}{3}] emitters at $z \sim 0.8$, and 22 cool dwarfs, as detailed below.
The photometric properties of the observed candidates are listed in Table \ref{tab:obsjournal}.
Table \ref{tab:nir_photometry} lists the objects detected in the near-IR bands.

Figures \ref{fig:spectra1} -- \ref{fig:spectra3} present 28 new quasars we identified at $5.7 \le z \le 7.0$.
Their spectroscopic properties are presented in the first section of Table \ref{tab:spectroscopy}.
The seven quasars at $z > 6.6$ are observed as $z$-dropouts on the HSC images, while the other objects are $i$-dropouts.
The quasars 
have broad Ly$\alpha$ lines, blue rest-UV continua, and/or sharp continuum breaks just blueward of Ly$\alpha$, the properties which are characteristic of high-$z$ quasars.
Following the previous papers, we classified the objects with very luminous ($L > 10^{43}$ erg s$^{-1}$) and narrow (full width at half maximum [FWHM] $<$ 500 km s$^{-1}$) 
Ly$\alpha$ emission as possible quasars (see the discussion below).
These objects 
tend to be found at the faintest magnitudes of our selection \citep{p4}, and have been missed in past shallower surveys. 

Redshifts of the discovered quasars were determined from the Ly$\alpha$ lines, assuming that the observed line peaks correspond to the intrinsic Ly$\alpha$ wavelength 
(1216 \AA\ in the rest frame).
This assumption is not always correct, due to the strong \ion{H}{1} absorption by the neutral IGM.
More accurate redshifts require observations of other emission lines, such as \ion{Mg}{2} $\lambda$2800 observed in the near-IR or [\ion{C}{2}] 158 $\mu$m accessible with ALMA.
When there is no clear Ly$\alpha$ lines, we obtained rough estimates of redshift from the wavelengths of the onset of the \citet{gunn65} trough.
Therefore the redshifts presented here are only approximate, with the uncertainties up to ${\Delta}z \sim 0.1$.

Absolute magnitudes ($M_{1450}$) and Ly$\alpha$ line properties of the quasars were measured as follows.
For every object, we defined a continuum window at wavelengths relatively free from strong sky emission lines, and extrapolated the measured continuum flux to estimate $M_{1450}$. 
A power-law continuum model with a slope $\alpha = -1.5$ ($F_{\lambda} \propto \lambda^{\alpha}$; e.g., \cite{vandenberk01}) was assumed.
Since the continuum windows fall in the range of $\lambda_{\rm rest}$ = 1220 -- 1350 \AA, which are close to $\lambda_{\rm rest}$ = 1450 \AA, 
these measurements are not sensitive to the exact value of $\alpha$.
The Ly$\alpha$ properties (luminosity, FWHM, and rest-frame equivalent width [EW]) of a quasar with relatively weak continuum emission, such as $J084456.62+022640.5$, 
was measured with a local continuum 
defined on the red side of Ly$\alpha$. 
For the remaining objects with strong continuum, we measured the properties of the broad Ly$\alpha$ + \ion{N}{5} $\lambda$1240 complex, with a local continuum defined by the above power-law model.
The resultant line properties are summarized in Table \ref{tab:spectroscopy}.

$J235646.33+001747.3$ has a Ly$\alpha$ line peak at 9740 \AA, which corresponds to a redshift $z = 7.01$. 
But this redshift may be an overestimate, as the intrinsic Ly$\alpha$ peak may be at a shorter wavelength and absorbed by the IGM.
The flux spike at $\sim$9800 \AA in the $J011257.84+011042.4$ spectrum is likely due to residual of a strong sky emission line.
$J151248.71+442217.5$ has a relatively red continuum, and indeed the HSC magnitudes from the latest DR 
($i - z = 1.5$ and $z - y = 0.5$) indicates a Bayesian quasar probability of $P_{\rm Q}^{\rm B} = 0.000$ (see Table \ref{tab:obsjournal}); this object was selected from an older DR,
which happened to give a higher $P_{\rm Q}^{\rm B}$.

Only three quasars are detected in the near-IR bands (see Table \ref{tab:nir_photometry}) and, interestingly, two of them ($J113753.64+004509.7$ and $J142307.21-022519.0$) 
apparently lack strong Ly$\alpha$ in emission.
We re-examined all the 75 broad-line quasars we discovered so far, and found that out of the eight quasars with near-IR detection, five are such weak-line quasars.
This is significantly higher than the average fraction of weak-line quasars among all 75 quasars ($\sim$20 \%; Y. Matsuoka et al., in prep.), 
or among more luminous quasars in the literature \citep[$\sim$10 \%][]{diamondstanic09, banados16}.
This may suggest a link between observed Ly$\alpha$ weakness and the continuum shape at longer wavelength.
We defer further analysis of this topic to a future paper.

Figure \ref{fig:spectra4} presents 7 high-$z$ objects with 
no or weak Ly$\alpha$ emission line, 
which are most likely galaxies at $z \sim 6$.
Redshifts of these galaxies were estimated from the observed wavelengths of Ly$\alpha$, the interstellar absorption lines of \ion{Si}{2} $\lambda$1260, 
\ion{Si}{2} $\lambda$1304, \ion{C}{2} $\lambda$1335, and/or the onset of the \citet{gunn65} trough. 
Due to the limited S/N of the spectra, the estimated redshifts should be regarded as only approximate
(${\Delta}z \la 0.1$).
The absolute magnitudes and Ly$\alpha$ properties were measured in the same way as for quasars, except that we assumed a continuum slope of
$\beta = -2.0$ \citep[$F_{\lambda} \propto \lambda^{\beta}$;][]{stanway05}.

Our quasar survey explores the luminosities where quasars and galaxies have comparable number densities \citep{p5}, and hence contamination of galaxies is inevitable.
Given the limited wavelength coverage of our spectra ($\lambda_{\rm rest} \sim 1200 - 1500$ \AA), high-$z$ objects without broad Ly$\alpha$ emission are difficult to classify unambiguously into quasars or galaxies.
Therefore the above quasar/galaxy classification is not perfect, and it may change in the future, with new data providing additional S/N or wavelength coverage.


In addition to the above high-$z$ objects, we found 10 [\ion{O}{3}] emitters at $z \sim 0.8$, as displayed in Figure \ref{fig:spectra5}.
Their strong [\ion{O}{3}] lines contribute significantly to the HSC $z$-band magnitude, and thus mimic colors of $z \sim 6 $ quasars.
We measured the properties of the H$\gamma$, H$\beta$, [\ion{O}{3}] $\lambda$4959 and $\lambda$5007 emission lines, as listed in Table \ref{tab:spectroscopy}.
Since these galaxies have very weak continuum, we estimated the continuum levels by summing up all available pixels after masking the above emission lines. 
As we discussed in \citet{p2}, their extremely high [\ion{O}{3}] $\lambda$5007/H$\beta$ ratios may indicate that these are galaxies with sub-solar metallicity and 
high ionization state of the interstellar medium \citep[e.g.,][]{kewley16}, and/or contribution from an active galactic nucleus (AGN).

The remaining 22 objects presented in Figures \ref{fig:spectra6} -- \ref{fig:spectra7} were 
found to be Galactic cool dwarfs (low-mass stars and brown dwarfs).
Their rough spectral classes were estimated by fitting the spectral standard templates of M4- to T8-type dwarfs, taken from the SpeX Prism Spectral Library \citep{burgasser14, skrzypek15}, 
to the observed spectra at $\lambda = 7500 - 10200$ \AA.
The results are summarized in Table \ref{tab:bdtypes} and plotted in the figures.
Due to the low S/N and limited wavelength coverage of the spectra, the classifications presented here are rather uncertain, and should be regarded as only approximate.

\begin{figure}
\epsscale{1.2}
\plotone{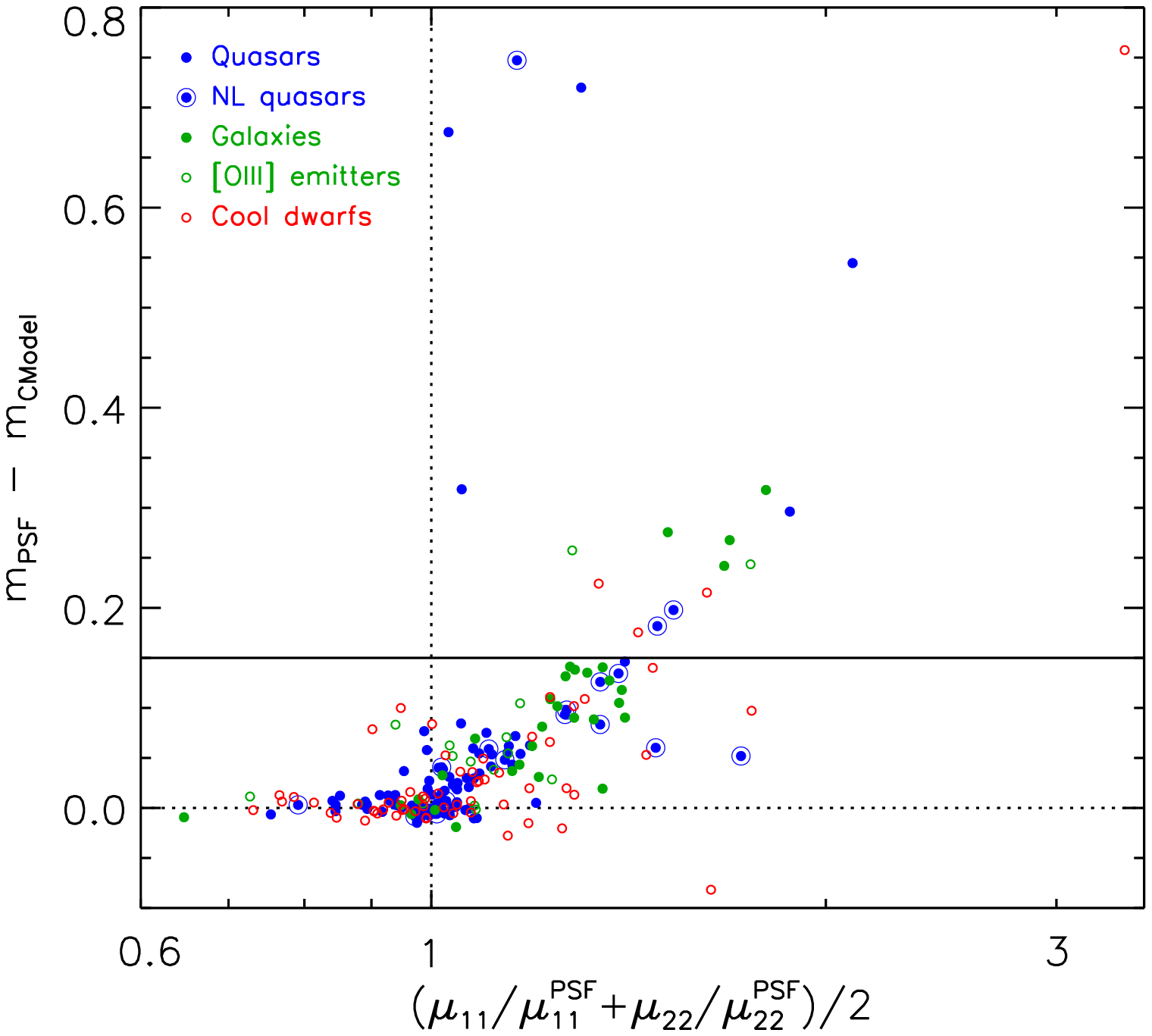}
\caption{Shape indicators of the quasars (blue dots, including previously-known quasars in the HSC survey footprint), 
narrow-line quasars (blue dots with larger circles), high-$z$ galaxies (green dots),
[\ion{O}{3}] emitters (green open circles), and Galactic cool dwarfs (red open circles).
For each object, we calculated these quantities in either of the $z$ or $y$ band with higher S/N.
Ideal point sources meet $m_{\rm PSF} = m_{\rm CModel}$ and $\mu_{11}/\mu_{11}^{\rm PSF} = \mu_{22}/\mu_{22}^{\rm PSF} = 1$, which are
represented by the two dotted lines.
The solid line represents our criterion of point source selection ($m_{\rm PSF} - m_{\rm CModel} < 0.15$).
\label{fig:shape}
}
\end{figure}

Figure \ref{fig:shape} presents shape indicators of the HSC objects with spectroscopic identification from our survey.
The horizontal axis uses second-order adaptive moments in two image directions \citep[$\mu_{11}$ and $\mu_{22}$;][]{hirata03}, which should be equal to
those of PSF ($\mu_{11}^{\rm PSF}$ and $\mu_{22}^{\rm PSF}$) for ideal point sources.
The figure shows that quasars with narrow Ly$\alpha$ lines are slightly more extended than the other quasars, which indicates contribution of stellar emission
(see the discussion further below).
We could eliminate some of the galaxy contamination by a stricter cut of point source selection (e.g., $m_{\rm PSF} - m_{\rm CModel} < 0.10$) than currently used, 
but that would also reject those potentially important population of narrow-line objects.
On the other hand, the figure suggests that the adoptive moments ($\mu_{11}$ and $\mu_{22}$) may be less affected by catastrophic measurement errors,
and hence may be a better shape indicator, than the magnitude difference ($m_{\rm PSF} - m_{\rm CModel}$);
this option will be considered in future candidate selection.

\begin{figure}
\epsscale{1.2}
\plotone{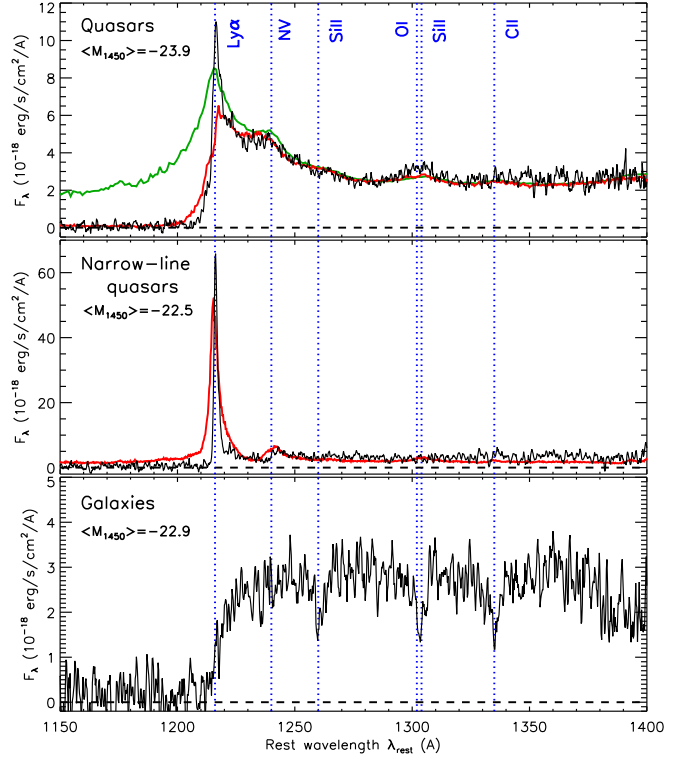}
\caption{Rest-frame composite spectra of all 75 broad-line quasars (top), 18 narrow-line quasars (middle), and 31 galaxies (bottom) discovered by our survey.
The median absolute magnitude $\langle M_{1450} \rangle$ of each sample is reported at the top left corner of each panel.
The red and green lines in the top panel represent the arbitrarily-scaled composite spectrum of luminous high-$z$ quasars from Pan-STARRS1 \citep{banados16}
and that of SDSS low-$z$ quasars \citep{vandenberk01}, respectively,
while the red line in the middle panel represent that of narrow-line quasar candidates in the SDSS \citep{alexandroff13}.
The blue dotted lines mark the wavelengths of Ly$\alpha$, \ion{N}{5} $\lambda$1240, 
\ion{Si}{2} $\lambda$1260, \ion{O}{1} $\lambda$1302, \ion{Si}{2} $\lambda$1304, and \ion{C}{2} $\lambda$1335.
\label{fig:stack}}
\end{figure}

Figure \ref{fig:stack} presents the composite spectra of all 75 broad-line quasars, 18 narrow-line quasars, and 31 galaxies discovered by our survey.
These spectra were generated by converting the individual spectra to the rest frame and normalizing to $M_{1450} = -23$ mag, and then median-stacking.
Thus the individual objects have equal weights in the stacking, regardless of the brightness or spectral S/N.
The overall shape of the quasar spectrum is similar to that of the local quasar composite \citep{vandenberk01},
except for the narrower Ly$\alpha$ line and the IGM absorption.
Our composite spectrum is also similar in shape to a composite spectrum of high-$z$ luminous quasars from the Panoramic Survey Telescope \& Rapid Response System 1 
\citep[Pan-STARRS1;][]{chambers06} survey \citep{banados16}, again except for the narrower Ly$\alpha$ line, which may reflect the \citet{baldwin77} effect.
While a clear sign of the quasar near-zone effect is visible just blueward of Ly$\alpha$, it should be noted that the redshifts of most of our quasars
have been determined with Ly$\alpha$, and are thus not very accurate.
A detailed analysis of emission and absorption profiles around Ly$\alpha$, including the effect of IGM damping-wing absorption, must wait for
accurate measurements of systemic redshifts via other emission lines. 

The composite spectrum of the narrow-line quasars is characterized by the strong and asymmetric Ly$\alpha$ line.
The rest-frame EW of this line is 29 $\pm$ 2 \AA\ in the composite, while it ranges up to $\sim$500 \AA\ in the individual spectra.
The spectrum shows a P Cygni-like profile around 1240 \AA, which is likely due to \ion{N}{5} $\lambda$1240 emission line and an associated mini broad absorption line system.
This spectral feature, along with the very luminous Ly$\alpha$ line usually associated with AGN \citep[$L_{\rm Ly\alpha} >10^{43}$ erg s$^{-1}$;][]{konno16} 
and the absence of interstellar absorption in the continuum, strongly suggest that these objects are narrow-line quasars \citep[see also the discussion in][]{p2}.
Compared with narrow-line quasar candidates in SDSS \citep{alexandroff13}, these high-$z$ objects have apparently narrower Ly$\alpha$,
indeed many are spectrally unresolved 
(most of the narrow-line quasars were observed with Subaru/FOCAS, which has a resolving power of
$\sim$250 km s$^{-1}$ with our observing mode).
This may suggest a significant contribution from the host galaxies, thus these objects may be composites of quasars (or AGNs) and star-forming galaxies.

We note that a color selection of high-$z$ quasars (including a more sophisticated Bayesian selection we used) is generally more sensitive to objects 
with stronger Ly$\alpha$ lines.
Thus the composite spectra may not represent the whole populations of broad- and narrow-line quasars that reside in the high-$z$ universe.
On the other hand, we confirmed in a previous work \citep[LF paper;][]{p5} that our quasar selection is fairly complete, as long as high-$z$ quasars
have a similar distribution of intrinsic spectral shapes to low-$z$ SDSS quasars.

Finally, the galaxy spectrum in Figure \ref{fig:stack} shows strong absorption lines, with rest-frame EWs of $1.7 \pm 0.8$ \AA, $1.8 \pm 0.7$ \AA, and $2.2 \pm 1.6$ \AA\
for \ion{Si}{2} $\lambda$1260, \ion{Si}{2} $\lambda$1304, and \ion{C}{2} $\lambda$1335, respectively.
These are broadly consistent with the values measured in a composite spectrum of $z \sim 3$ Lyman break galaxies \citep{shapley03}. 
The galaxies discovered by our survey are very luminous, with $M_{1450} \simeq -22$ to $-25$ mag, well above the break magnitude of the galaxy LF
at $z = 6$ \citep[$M_{1450} \simeq -21$ mag;][]{ono18}. 

\begin{figure}
\epsscale{1.2}
\plotone{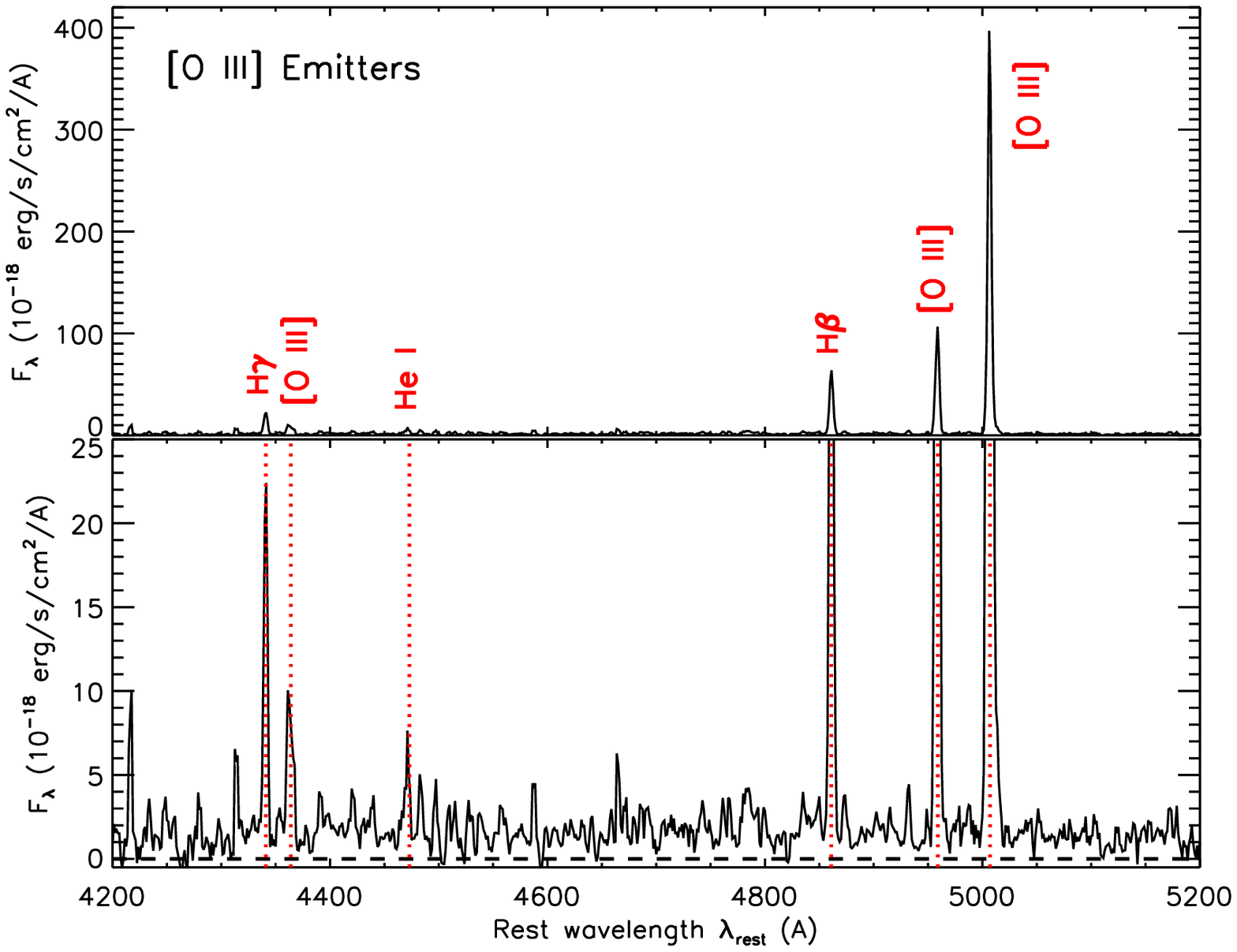}
\caption{Rest-frame composite spectrum of all 16 [\ion{O}{3}] emitters at $z \sim 0.8$ over the full (top) and lower (bottom) flux scales.
The red dotted lines mark the wavelengths of H$\gamma$, [\ion{O}{3}] $\lambda$4364, \ion{He}{1} $\lambda$4473, H$\beta$, [\ion{O}{3}] $\lambda$4959 and $\lambda$5007.
\label{fig:stack_o3}}
\end{figure}

We also present a composite spectrum of the 16 [\ion{O}{3}] emitters at $z \sim 0.8$ in Figure \ref{fig:stack_o3}.
This spectrum was created by converting the individual spectra to the rest frame and normalizing to the continuum flux of 
$10^{18}$ erg s$^{-1}$ cm$^{-2}$ \AA$^{-1}$, and then median-stacking.
In addition to H$\gamma$, H$\beta$, [\ion{O}{3}] $\lambda$4959 and $\lambda$5007, we detected [\ion{O}{3}] $\lambda$4363 and \ion{He}{1} $\lambda$4473 emission;
these two lines are weak and undetected in the individual spectra.
The rest-frame EWs of the six emission lines are 53 $\pm$ 3 \AA\ (H$\gamma$), 35 $\pm$ 3 \AA\ ([\ion{O}{3}] $\lambda$4363), 21 $\pm$ 3 \AA\ (\ion{He}{1} $\lambda$4473), 
165 $\pm$ 7 \AA\ (H$\beta$), 264 $\pm$ 10 \AA\ ([\ion{O}{3}] $\lambda$4959), 1030 $\pm$ 40 \AA\ ([\ion{O}{3}] $\lambda$5007).
All lines are spectrally unresolved.




The HSC-SSP survey has completed observations on more than 80 \% of the planned 300 nights, and we are making steady progress on our high-$z$ quasar survey.
We plan to continue our follow-up spectroscopy in the next few years, and will report new discoveries, along with measurements of the properties of individual objects
with multi-wavelength observations.
The quasar sample thus established will provide a more accurate quasar LF at $z \ge 6$ than is currently available \citep[e.g.,][]{p5, wang18}.
In a few years, we also aim to start a large SSP survey with the Prime Focus Spectrograph, a new wide-field multi-object spectrograph for the Subaru Telescope
under development \citep{takada14}.
This will enable us to take spectra of a significant number of HSC objects, including quasar candidates at all redshifts.

\startlongtable
\begin{deluxetable*}{lcccccc}
\tablecaption{Spectroscopic Properties\label{tab:spectroscopy}}
\tablehead{
\colhead{Name} & \colhead{Redshift} & 
\colhead{$M_{1450}$} & \colhead{Line} & 
\colhead{EW$_{\rm rest}$} & \colhead{FWHM} & 
\colhead{log $L_{\rm line}$}\\
\colhead{} & \colhead{} & 
\colhead{(mag)} & \colhead{} & 
\colhead{(\AA)} & \colhead{(km s$^{-1}$)} & 
\colhead{($L_{\rm line}$ in erg s$^{-1}$)}
} 
\startdata
\multicolumn{7}{c}{Quasars}\\\hline
$J235646.33+001747.3$ &  7.01 &   $-25.31 \pm 0.04$              & \nodata &               \nodata &               \nodata &               \nodata \\   
$J160953.03+532821.0$ &  6.92 &   $-22.75 \pm 1.67$           & Ly$\alpha$ &     210 $\pm$ 320 &     3600 $\pm$ 1300 &    44.31 $\pm$ 0.07 \\  
$J011257.84+011042.4$ &  6.82 &   $-24.07 \pm 0.35$           & Ly$\alpha$ &     15 $\pm$ 7 &     1500 $\pm$ 400     &    43.71 $\pm$ 0.14 \\   
$J161207.12+555919.2^*$ &  6.78 &   $-23.02 \pm 0.32$           & Ly$\alpha$ &     12 $\pm$ 8 &     560 $\pm$ 270       &    43.15 $\pm$ 0.14 \\  
$J134400.87+012827.8$ &  6.72 &   $-23.46 \pm 0.15$           & Ly$\alpha$ &     73 $\pm$ 12 &     9000 $\pm$ 1100 &    44.13 $\pm$ 0.04 \\ 
$J000142.54+000057.5$ &  6.69 &   $-24.49 \pm 0.59$           & Ly$\alpha$ &     11 $\pm$ 15 &     1600 $\pm$ 1000 &    43.72 $\pm$ 0.59 \\   
$J123137.77+005230.3$ &  6.69 &   $-24.39 \pm 0.09$           & Ly$\alpha$ &     27 $\pm$ 3 &     2400 $\pm$ 900     &    44.09 $\pm$ 0.03 \\ 
$J135012.04-002705.2$ &  6.49 &   $-24.38 \pm 0.19$           & Ly$\alpha$ &     56 $\pm$ 11 &     620 $\pm$ 200      &    44.39 $\pm$ 0.04 \\  
$J084456.62+022640.5^*$ &  6.40 &   $-21.57 \pm 0.47$         & Ly$\alpha$ &     280 $\pm$ 160 &         $<$ 230          &    44.05 $\pm$ 0.01 \\ 
$J113753.64+004509.7$ &  6.4 &    $-24.20 \pm 0.13$           & Ly$\alpha$ &     22 $\pm$ 6 &     11000 $\pm$ 3000  &   43.91 $\pm$ 0.11 \\  
$J152555.79+430324.0$ &  6.27 &   $-23.61 \pm 0.06$           & Ly$\alpha$ &     35 $\pm$ 4 &     340 $\pm$ 740      &    43.88 $\pm$ 0.04 \\  
                                       &       &                                              & NV &               18 $\pm$ 2 &     2900 $\pm$ 1600 &    43.60 $\pm$ 0.04 \\  
$J151248.71+442217.5$ &  6.19 &   $-22.07 \pm 0.04$           & Ly$\alpha$ &     28 $\pm$ 2 &     3700 $\pm$ 200 &    43.18 $\pm$ 0.03 \\ 
                                      &       &                                              & NV            &     11 $\pm$ 2 &     2800 $\pm$ 300 &    42.77 $\pm$ 0.07 \\ 
$J225520.78+050343.3$ &  6.18 &   $-24.43 \pm 0.02$           & Ly$\alpha$ &     30 $\pm$ 1 &     6200 $\pm$ 700 &    44.14 $\pm$ 0.02 \\  
$J134733.69-015750.6$ &  6.15 &   $-24.73 \pm 0.02$           & Ly$\alpha$ &     65 $\pm$ 2 &     1480 $\pm$ 10    &    44.59 $\pm$ 0.01 \\ 
$J144823.33+433305.9$ &  6.14 &   $-24.36 \pm 0.04$           & Ly$\alpha$ &     26 $\pm$ 2 &     7200 $\pm$ 1900 &    44.05 $\pm$ 0.03 \\  
$J000133.30+000605.4$ &  6.13 &   $-23.72 \pm 0.06$           & Ly$\alpha$ &     11 $\pm$ 1 &     2500 $\pm$ 1100 &    43.44 $\pm$ 0.04 \\ 
$J151657.87+422852.9$ &  6.13 &   $-24.35 \pm 0.01$           & Ly$\alpha$ &     17 $\pm$ 1 &     9500 $\pm$ 2100 &    43.85 $\pm$ 0.03 \\  
$J125437.08-001410.7^*$ &  6.13 &   $-20.91 \pm 0.32$         & Ly$\alpha$ &     470 $\pm$ 160 &     310 $\pm$ 20 &    44.03 $\pm$ 0.01 \\ 
$J000445.81-004944.3$ &  6.10 &   $-23.90 \pm 0.06$           & Ly$\alpha$ &     53 $\pm$ 4 &     2400 $\pm$ 900 &    44.18 $\pm$ 0.02 \\  
$J093543.32-011033.3^*$ &  6.08 &   $-21.97 \pm 0.18$        & Ly$\alpha$ &     410 $\pm$ 170 &          $<$ 230    &    44.12 $\pm$ 0.01 \\ 
$J010603.68-003015.2$ &  6.06 &   $-23.53 \pm 0.05$           & Ly$\alpha$ &     46 $\pm$ 3 &     3300 $\pm$ 600 &    43.97 $\pm$ 0.02 \\  
$J142611.33-012822.8$ &  6.01 &   $-23.75 \pm 0.10$           & Ly$\alpha$ &     11 $\pm$ 4 &     4100 $\pm$ 300 &    43.45 $\pm$ 0.13 \\  
$J121049.13+013426.7$ &  5.97 &   $-22.60 \pm 0.07$       & \nodata &    \nodata &     \nodata &    \nodata \\ 
$J093000.85+005738.4$ &  5.92 &   $-24.91 \pm 0.05$           & Ly$\alpha$ &     24 $\pm$ 4 &     13900 $\pm$ 400 &    44.22 $\pm$ 0.07 \\  
$J161819.77+552654.0$ &  5.91 &   $-24.26 \pm 0.09$           & Ly$\alpha$ &     32 $\pm$ 5 &     8100 $\pm$ 3300 &    44.10 $\pm$ 0.06 \\   
$J142307.21-022519.0$ &  5.9 &   $-24.26 \pm 0.13$           & Ly$\alpha$ &     26 $\pm$ 7   &     7400 $\pm$ 3300 &    44.01 $\pm$ 0.10 \\  
$J023858.09-031845.4$ &  5.83 &   $-23.94 \pm 0.03$           & Ly$\alpha$ &     45 $\pm$ 3 &     14700 $\pm$ 100 &    44.11 $\pm$ 0.02 \\  
$J113218.15+003800.1$ &  5.66 &   $-23.18 \pm 0.05$           & Ly$\alpha$ &     24 $\pm$ 4 &     650 $\pm$ 130 &      43.34 $\pm$ 0.05 \\ 
\hline\multicolumn{7}{c}{Galaxies}\\\hline
$J135348.55-001026.5$ &  6.2 &   $-24.76 \pm 0.02$              & \nodata &               \nodata &               \nodata &               \nodata \\  
$J144216.08+423632.5$ &  6.0 &   $-22.93 \pm 0.07$              & \nodata &               \nodata &               \nodata &               \nodata \\ 
$J092117.65+030521.5$ &  6.0 &   $-22.76 \pm 0.16$              & \nodata &               \nodata &               \nodata &               \nodata \\  
$J115755.51-001356.2$ &  5.9 &   $-22.98 \pm 0.08$              & \nodata &               \nodata &               \nodata &               \nodata \\  
$J123841.97-011738.8$ &  5.8 &   $-23.37 \pm 0.04$              & \nodata &               \nodata &               \nodata &               \nodata \\ 
$J162657.22+431133.0$ &  5.8 &   $-22.78 \pm 0.04$              & \nodata &               \nodata &               \nodata &               \nodata \\ 
$J020649.98-020618.2$ &  5.72 &   $-23.83 \pm 0.03$           & Ly$\alpha$ &     1.1 $\pm$ 0.2 &     250 $\pm$ 170 &    42.39 $\pm$ 0.08 \\ 
\hline\multicolumn{7}{c}{[\ion{O}{3}] Emitters}\\\hline
$J090017.67-014655.6$ & 0.921 & \nodata & H$\beta$                            &     16 $\pm$ 1     &             $<$ 190 &    40.99 $\pm$ 0.03 \\ 
                      &       &               \nodata & [\ion{O}{3}] $\lambda$4959 &     8.7 $\pm$ 0.6 &             $<$ 190 &    40.71 $\pm$ 0.03 \\ 
                      &       &               \nodata & [\ion{O}{3}] $\lambda$5007 &     31 $\pm$ 1    &             $<$ 190 &    41.26 $\pm$ 0.02 \\ 
$J015519.63-005814.7$ & 0.872 & \nodata         & H$\gamma$            &     6.5 $\pm$ 0.6 &             $<$ 190 &    40.37 $\pm$ 0.04 \\  
                      &       &               \nodata             & H$\beta$                &     28 $\pm$ 1    &     180 $\pm$ 20 &    41.00 $\pm$ 0.02 \\  
                      &       &               \nodata & [\ion{O}{3}] $\lambda$4959 &     33 $\pm$ 1    &     150 $\pm$ 40 &    41.07 $\pm$ 0.01 \\  
                      &       &               \nodata & [\ion{O}{3}] $\lambda$5007 &     71 $\pm$ 6    &              $<$ 190 &    41.41 $\pm$ 0.03 \\  
$J144221.95-013258.0$ & 0.823       &               \nodata             & H$\beta$ &     200 $\pm$ 80 &             $<$ 230 &    40.81 $\pm$ 0.06 \\ 
                      &       &               \nodata & [OIII] $\lambda$4959 &     290 $\pm$ 110 &             $<$ 230 &    40.97 $\pm$ 0.05 \\ 
                      &       &               \nodata & [OIII] $\lambda$5007 &     1600 $\pm$ 600 &             $<$ 230 &    41.70 $\pm$ 0.01 \\ 
$J115946.07+014023.7$ & 0.790 &               \nodata            & H$\gamma$ &     33 $\pm$ 7 &             $<$ 190 &    41.24 $\pm$ 0.07 \\  
                      &       &               \nodata             & H$\beta$ &     150 $\pm$ 20 &             $<$ 190 &    41.89 $\pm$ 0.02 \\  
                      &       &               \nodata & [OIII] $\lambda$4959 &     350 $\pm$ 40 &             $<$ 190 &    42.26 $\pm$ 0.01 \\  
                      &       &               \nodata & [OIII] $\lambda$5007 &     1000 $\pm$ 100 &             $<$ 190 &    42.73 $\pm$ 0.01 \\  
$J141332.93+014350.4$ & 0.784 &               \nodata             & H$\beta$ &                 $>$ 1500 &             $<$ 230 &    40.99 $\pm$ 0.05 \\ 
                      &       &               \nodata & [OIII] $\lambda$4959 &                 $>$ 2800 &             $<$ 230 &    41.28 $\pm$ 0.02 \\ 
                      &       &               \nodata & [OIII] $\lambda$5007 &                 $>$ 9700 &             $<$ 230 &    41.82 $\pm$ 0.01 \\ 
$J094209.69-020302.1$ & 0.782 & \nodata             & H$\beta$                 &     71 $\pm$ 5   &             $<$ 190 &    41.61 $\pm$ 0.02 \\ 
                      &       &               \nodata & [\ion{O}{3}] $\lambda$4959 &     110 $\pm$ 7  &             $<$ 190 &    41.79 $\pm$ 0.02 \\ 
                      &       &               \nodata & [\ion{O}{3}] $\lambda$5007 &     380 $\pm$ 20 &             $<$ 190 &    42.34 $\pm$ 0.01 \\ 
$J225124.84-010911.0$ & 0.777 & \nodata         & H$\gamma$            &     95 $\pm$ 11   &             $<$ 190 &    41.00 $\pm$ 0.04 \\ 
                      &       &               \nodata             & H$\beta$                 &     210 $\pm$ 20 &             $<$ 190 &    41.34 $\pm$ 0.02 \\ 
                      &       &               \nodata & [\ion{O}{3}] $\lambda$4959 &     420 $\pm$ 40 &             $<$ 190 &    41.65 $\pm$ 0.01 \\ 
                      &       &               \nodata & [\ion{O}{3}] $\lambda$5007 &     1300 $\pm$ 100 &         $<$ 190 &    42.13 $\pm$ 0.01 \\ 
$J093831.54-002523.4$ & 0.777 &               \nodata            & H$\gamma$ &     390 $\pm$ 270 &             $<$ 230 &    40.64 $\pm$ 0.07 \\ 
                      &       &               \nodata             & H$\beta$ &     600 $\pm$ 410 &             $<$ 230 &    40.83 $\pm$ 0.04 \\ 
                      &       &               \nodata & [OIII] $\lambda$4959 &     1300 $\pm$ 900 &             $<$ 230 &    41.18 $\pm$ 0.02 \\ 
                      &       &               \nodata & [OIII] $\lambda$5007 &     3800 $\pm$ 2600 &             $<$ 230 &    41.63 $\pm$ 0.01 \\ 
$J135049.56+013726.6$ & 0.770 &               \nodata            & H$\gamma$ &                 $>$ 110 &             $<$ 230 &    40.43 $\pm$ 0.06 \\ 
                      &       &               \nodata             & H$\beta$ &                 $>$ 350 &             $<$ 230       &    40.93 $\pm$ 0.02 \\ 
                      &       &               \nodata & [OIII] $\lambda$4959 &                 $>$ 470 &             $<$ 230 &    41.06 $\pm$ 0.05 \\ 
                      &       &               \nodata & [OIII] $\lambda$5007 &                 $>$ 2100 &             $<$ 230 &    41.70 $\pm$ 0.01 \\ 
$J150602.11+415317.5$ & 0.764 & \nodata        & H$\gamma$            &     103 $\pm$ 4    &             $<$ 190 &    41.30 $\pm$ 0.01 \\ 
                      &       &               \nodata             & H$\beta$                &     249 $\pm$ 8    &             $<$ 190 &    41.68 $\pm$ 0.01 \\ 
                      &       &               \nodata & [\ion{O}{3}] $\lambda$4959 &     540 $\pm$ 20  &             $<$ 190 &    42.02 $\pm$ 0.01 \\ 
                      &       &               \nodata & [\ion{O}{3}] $\lambda$5007 &     1600 $\pm$ 100 &          $<$ 190 &    42.50 $\pm$ 0.01 \\ 
\enddata
\tablecomments{The asterisks after the object names indicate the possible quasars with narrow Ly$\alpha$ emission.
Upper limits of line EWs are placed at $3\sigma$ significance, for the objects without continuum detection.}
\end{deluxetable*}

\begin{deluxetable}{cc}
\tablecaption{Spectral classes of the cool dwarfs\label{tab:bdtypes}}
\tablehead{
\colhead{Name} & \colhead{Class}
} 
\startdata
$J011152.80+013211.4$ & L4 \\
$J015548.62-061737.9$ & M4 \\
$J020421.81-024421.1$ & L9 \\
$J022144.40-053008.6$ & T2 \\
$J085230.35+025410.0$ & M6 \\
$J085818.99+040927.6$ & M7 \\
$J124505.57+010550.2$ & M9 \\
$J135913.21+002740.0$ & T8 \\
$J142531.82-021606.5$ & T0 \\
$J145853.06+015031.8$ & L9 \\
$J151812.04+440829.0$ & T1 \\
$J153032.76+435615.5$ & L8 \\
$J154605.67+425306.1$ & T7 \\
$J161042.47+554203.4$ & T0\\
$J164226.30+430749.7$ & L1 \\
$J223827.66+053246.5$ & M4 \\
$J224157.52+055912.6$ & M5 \\
$J225132.39-011601.1$ & M5 \\
$J225337.55+051440.1$ & M7 \\
$J225513.58+044226.8$ & M5 \\
$J230811.42+023931.3$ & M7 \\
$J231344.68-003931.3$ & L1 \\
\enddata
\tablecomments{These classification should be regarded as only approximate; see the text.}
\end{deluxetable}

\acknowledgments

We are grateful to the referee for his/her useful comments to improve this paper.8
We thank Rachael Alexandroff for kindly sharing with us the electronic data of the narrow-line quasar composite spectrum in Figure \ref{fig:stack}. 

This work is based on data collected at the Subaru Telescope, which is operated by the National Astronomical Observatory of Japan (NAOJ).
We appreciate the staff members of the telescope for their support during our FOCAS observations.
The data analysis was in part carried out on the open use data analysis computer system at the Astronomy Data Center of NAOJ.

This work is also based on observations made with the Gran Telescopio Canarias (GTC), installed at the Spanish Observatorio del Roque de los Muchachos 
of the Instituto de Astrof\'{i}sica de Canarias, on the island of La Palma.
We thank Stefan Geier and other support astronomers for their help during preparation and execution of our observing program.

Y. M. was supported by the Japan Society for the Promotion of Science (JSPS) KAKENHI Grant No. JP17H04830 and the Mitsubishi Foundation grant No. 30140.
K. I. acknowledges support by the Spanish MINECO under grant AYA2016-76012-C3-1-P and MDM-2014-0369 of ICCUB (Unidad de Excelencia 'Mar\'ia de Maeztu')

The Hyper Suprime-Cam (HSC) collaboration includes the astronomical
communities of Japan and Taiwan, and Princeton University.  The HSC
instrumentation and software were developed by NAOJ, the Kavli Institute for the
Physics and Mathematics of the Universe (Kavli IPMU), the University
of Tokyo, the High Energy Accelerator Research Organization (KEK), the
Academia Sinica Institute for Astronomy and Astrophysics in Taiwan
(ASIAA), and Princeton University.  Funding was contributed by the FIRST 
program from Japanese Cabinet Office, the Ministry of Education, Culture, 
Sports, Science and Technology (MEXT), the Japan Society for the 
Promotion of Science (JSPS),  Japan Science and Technology Agency 
(JST),  the Toray Science  Foundation, NAOJ, Kavli IPMU, KEK, ASIAA,  
and Princeton University.

This paper makes use of software developed for the Large Synoptic Survey Telescope (LSST). We thank the LSST Project for 
making their code available as free software at http://dm.lsst.org.

The Pan-STARRS1 Surveys (PS1) have been made possible through contributions of the Institute for Astronomy, the University of Hawaii, the Pan-STARRS Project Office, the Max-Planck Society and its participating institutes, the Max Planck Institute for Astronomy, Heidelberg and the Max Planck Institute for Extraterrestrial Physics, Garching, The Johns Hopkins University, Durham University, the University of Edinburgh, Queen's University Belfast, the Harvard-Smithsonian Center for Astrophysics, the Las Cumbres Observatory Global Telescope Network Incorporated, the National Central University of Taiwan, the Space Telescope Science Institute, the National Aeronautics and Space Administration under Grant No. NNX08AR22G issued through the Planetary Science Division of the NASA Science Mission Directorate, the National Science Foundation under Grant No. AST-1238877, the University of Maryland, E\"{o}tv\"{o}s Lorand University (ELTE) and the Los Alamos National Laboratory.

IRAF is distributed by the National 
Optical Astronomy Observatory, which is operated by the Association of Universities for Research in Astronomy (AURA) under a cooperative agreement 
with the National Science Foundation.

\end{document}